\documentclass[11pt]{article}
\usepackage[totalwidth=460truept,totalheight=600truept]{geometry}
\usepackage{epsfig,latexsym,amssymb}

\def\theequation{\arabic{section}.\arabic{equation}}

\renewcommand{\theequation}{\thesection.\arabic{equation}}
\global\arraycolsep=1truept
\input{tcilatex}
\begin{document}

\hfill \hfill IFUP-TH 2006-11

\vskip 1.4truecm

\begin{center}
{\huge \textbf{Renormalization And Causality Violations}}

{\large \textbf{\vskip .1truecm}}

{\huge \textbf{\ In\ Classical Gravity}}

{\large \textbf{\vskip .1truecm}}

{\huge \textbf{Coupled With Quantum Matter}}

\vskip 1.5truecm

\textsl{Damiano Anselmi}

\textit{Dipartimento di Fisica ``Enrico Fermi'', Universit\`{a} di Pisa, }

\textit{Largo Pontecorvo 3, I-56127 Pisa, Italy, }

\textit{and INFN, Sezione di Pisa, Pisa, Italy}

damiano.anselmi@df.unipi.it
\end{center}

\vskip 2truecm

\begin{center}
\textbf{Abstract}
\end{center}

\bigskip

{\small I prove that classical gravity coupled with quantized matter can be
renormalized with a finite number of independent couplings, plus field
redefinitions, without introducing higher-derivative kinetic terms in the
gravitational sector, but adding vertices that couple the matter
stress-tensor with the Ricci tensor. The theory is called ``acausal
gravity'', because it predicts the violation of causality at high energies.
Renormalizability is proved by means of a map }$\mathcal{M}${\small \ that
relates acausal gravity with higher-derivative gravity. The causality
violations are governed by two parameters, }$a${\small \ and }$b${\small ,
that are mapped by }$\mathcal{M}${\small \ into higher-derivative couplings.
At the tree level causal prescriptions exist, but they are spoiled by the
one-loop corrections. Some ideas are inspired by the usual treatments of the
Abraham-Lorentz force in classical electrodynamics.}

\vskip 1truecm

\vfill\eject

\section{Introduction}

\setcounter{equation}{0}

The necessity of quantizing gravity is a debated issue. Bohr and Rosenfeld 
\cite{bohr} showed that a theory in which some fields are quantized and
others are not can violate some basic principles of quantum mechanics, for
example the indeterminacy principle. Rosenfeld \cite{rosenfeld} observed
that there is no direct evidence for the validity of such principles in
situations where the gravitational field is important. Feynman questioned
whether gravity must be quantized in his lectures on gravitation \cite
{feynman}. M\o ller \cite{moller} and Rosenfeld \cite{rosenfeld} gave a
specific suggestion to couple a one-half quantum and one-half classical
world, in the realm of quantum mechanics. They stated that the spacetime
geometry couples to the expectation value of the energy-momentum tensor,
calculated on the quantum state of the matter fields. Eppley and Hannah \cite
{eppley} showed that if matter is quantized, but gravity is classical, then,
assuming the ``Copenhagen'' interpretation of quantum mechanics, two
scenarios are given: if the gravitational interactions do not collapse the
wave-function, gravity can be used to propagate information at superluminal
velocity; if, on the other hand, gravity collapses the wave-function, then
either the uncertainty principle or energy-momentum conservation can be
violated. They also suggested an experiment to establish whether gravity is
quantum mechanical. Recently, Mattingly \cite{mattingly} questioned the
feasibility of any such experiment. Other arguments advocated to assert that
gravity needs to be quantized are weaker, because they are just based on the
analogy with the other interactions of nature. None of these observations
settle the debate, actually, since experiments are unable, at present, to
ensure that the gravitational interactions obey the indeterminacy principle
and causality at arbitrarily high energies.

A remarkable fact is that the Standard Model is ``ready'' for the coupling
with gravity, in the sense that the anomaly cancellations survive the
embedding in a curved background \cite{peskin}. Thus it is natural to
consider a partially quantized theory where the Standard Model is embedded
in external gravity, which is treated classically, and the pure-gravity
sector is described just by the Einstein action with a cosmological term.
For consistency, no higher-derivative gravitational kinetic terms should be
turned on by renormalization.

The investigation of classical gravity coupled with quantum field theory in
a curved background is an alternative way to search for new physics beyond
the Standard Model. A variety of problems can be treated exactly and
physical predictions can be derived. The results can also suggest new
experimental observations to determine whether gravity must be quantized or
not. Some predictions might hold also for quantum gravity, at least
qualitatively.

The main purposes of this paper are to:

\noindent 1) extend the M\o ller-Rosenfeld approach \cite{moller,rosenfeld}
to quantum field theory, formulating a minimum principle that generates the
field equations of a partially quantized theory,

\noindent 2) prove that classical gravity coupled with quantum matter is
renormalizable with a finite number of independent parameters, without
introducing higher-derivative kinetic terms in the gravitational sector;

\noindent 3) analyze the physical effects of renormalization in the
gravitational sector, such as the violation of causality at short distances.

The quantization of fields in curved space (see for example \cite
{birreldavis}) has motivated an enormous amount of work. An extension of the
M\o ller-Rosenfeld approach has been proposed by Schwinger and Keldysh \cite
{schwingerkeldysh}, in terms of the ``in-in'' expectation value of the
stress tensor, which is both real and causal. The approach formulated here
uses out-in expectation values, to make a more direct connection with the
standard formulation of quantum field theory. Nevertheless, the other
results of this paper do not depend in a crucial way on how the
stress-tensor expectation value is interpreted. In particular, with some
obvious modifications, properties 2) and 3) hold also in the
Schwinger-Keldysh framework.

The renormalizability of the partially quantized theory is proved applying a
theorem stating that a term quadratically proportional to the field
equations can be reabsorbed by a field redefinition to all orders. For the
investigation of this paper, such a theorem is rephrased by a map 
\begin{equation}
\mathcal{M}:S_{\mathrm{HD}}\rightarrow S_{\mathrm{AC}}  \label{themap}
\end{equation}
that relates a causal theory $S_{\mathrm{HD}}$ with instabilities, typically
due to higher-derivative (HD) kinetic terms, with an acausal theory $S_{%
\mathrm{AC}}$ without instabilities. Precisely, $S_{\mathrm{HD}}$ is
higher-derivative classical gravity coupled with quantum matter, whose
renormalization is straightforward. The matter fields circulating in the
loops generate the higher-derivative counterterms $R_{\mu \nu }R^{\mu \nu }$
and $R^{2}$, which are subtracted adding these same terms to the lagrangian,
multiplied by independent parameters $a$ and $b$. Instead, $S_{\mathrm{AC}}$
denotes classical Einstein gravity coupled with quantum matter. Its
renormalization is less trivial. The counterterms $R_{\mu \nu }R^{\mu \nu }$
and $R^{2}$ are subtracted by means of a field redefinition of the metric
tensor. The existence of such a field redefinition is obvious to the lowest
order. The map $\mathcal{M}$ ensures its existence to all orders. In
practice, the map $\mathcal{M}$ replaces the higher-derivative terms $R_{\mu
\nu }R^{\mu \nu }$ and $R^{2}$ by new vertices belonging to the matter
sector, that couple the matter stress tensor to the Ricci tensor, with
coupling constants $a$ and $b$.

A typical feature of higher derivative theories is that the field equations
admit unstable solutions. For a discussion in classical higher-derivative
gravity, see for example \cite{magnano}. The field redefinition provided by
the map $\mathcal{M}$ eliminates the unstable solutions. On the other hand,
the map $\mathcal{M}$ contains power series in momenta that can be resummed
exactly. The main outcome of the resummation is the violation of causality
at high energies. The causality violation is independent of the
interpretation of the stress-tensor expectation value. In particular, it is
present also in the Schwinger-Keldysh approach.

The correspondence between instabilities and causality violations is
inspired by an analogous correspondence that is usually learnt in connection
with the Abraham-Lorentz force of classical electrodynamics \cite{jackson}
and that has been applied also to higher-derivative gravity \cite
{bel,parkersimon}. The approach (\ref{themap}) is not equivalent to the ones
existing in the literature and is specifically designed to work efficiently
in combination with renormalization.

The map $\mathcal{M}$ is useful to relate the renormalization properties of $%
S_{\mathrm{HD}}$\ and $S_{\mathrm{AC}}$, but it is not just a change of
variables. The theories $S_{\mathrm{HD}}$\ and $S_{\mathrm{AC}}$ are
physically inequivalent, because the unstable solutions of $S_{\mathrm{HD}}$
are not solutions of $S_{\mathrm{AC}}$. The map $\mathcal{M}$ is used to
show that classical gravity coupled with quantum matter is predictive,
because it can be renormalized with a finite set of independent couplings,
plus field redefinitions, without introducing higher-derivative kinetic
terms in the gravitational sector.

Commonly \cite{mottola} the Planck scale is considered as the physical
cut-off which defines the extreme limit of validity of semi-classical
gravity and the attention is confined to predictions that involve energy
scales much greater than the Planck length. However, as long as there is no
definitive experimental evidence that gravity should be quantized, nor that
causality should hold at arbitrarily high energies, there is no compelling
reason to consider the model of this paper as an effective one. In such a
situation, it belongs to the duties of a theorist to investigate also the
consequences that follow from the assumption that the model is a fundamental
theory, valid at arbitrarily high energies. This attitude is also the most
efficient one to eventually uncover reasons to reject the assumption. As
mentioned above, in the acausal theory constructed here, certain power
series in the momenta can be resummed exactly, so it is compulsory to take
these resummations seriously and inquire about their physical meaning, if
any. What happens is that the stress tensor gets averaged in an acausal way,
because the average receives contributions also from the future light cone
and from spacelike separated points. The causality violations are
parametrized by $a$ and $b^{\prime }=-2(a+3b)$. At the tree level, there
exist causal prescriptions, if $a$ and $b^{\prime }$ are negative. However,
the radiative corrections spoil the causal prescriptions and produce
causality violations in any case.

The physical effects of the couplings $a$ and $b^{\prime }$ can be detected
also in causal situations. Experimental bounds on the values of $a$ and $%
b^{\prime }$ can be derived from the tests about the validity of Newton's
law at short distances.

The map $\mathcal{M}$ cannot be applied to quantum gravity, at least in a
straightforward way. This is a weakness of the model if gravity ultimately
needs to be quantized. It is a good feature of the model, instead, if
gravity does not need to be quantized. Still, the results of this paper
might inspire the search for appropriate generalizations of the map $%
\mathcal{M}$ to quantum gravity.

\bigskip

The paper is organized as follows. The minimization principle for fully and
partially quantized field theories is treated in section 2. The map $%
\mathcal{M}$ is studied in section 3 and worked out explicitly for gravity
in the quadratic approximation. A source term is then added to study the
physical effects. In section 4 the map $\mathcal{M}$ is used to prove the
renormalizability of the theory. Section 5 is devoted to the investigation
of causality violations and their relation with instabilities. Section 6
contains the conclusions.

\section{Minimum principles for fully and partially quantized field theories}

\setcounter{equation}{0}

According to the M\o ller-Rosenfeld approach \cite{moller,rosenfeld}, in
quantum mechanics classical gravity couples to the expectation value of the
energy-momentum tensor, calculated on the quantum state $\psi $ of the
matter fields. The Einstein equations read 
\begin{equation}
R_{\mu \nu }-\frac{1}{2}g_{\mu \nu }R=-\kappa ^{2}\left\langle \psi \left|
T_{\mu \nu }\right| \psi \right\rangle .  \label{moller}
\end{equation}
The generalization of this equation to classical gravity coupled with
quantized fields has been discussed by various authors in the literature. In
the Schwinger-Keldysh \cite{schwingerkeldysh} approach, the right-hand side
of (\ref{moller}) is replaced with the ``in-in'' expectation value of the
stress tensor, so it is both real and causal. Functional methods for the
calculation of in-in expectation values have been developed \cite
{jordan,woodard}. It is important to observe that the renormalization
structure does not depend on the interpretation of the right-hand side of (%
\ref{moller}). In particular, the counterterms $R_{\mu \nu }R^{\mu \nu }$
and $R^{2}$ calculated in the Schwinger-Kleydish approach are identical to
those calculated in the usual approach \cite{woodard}. The causality
violations discussed here, which are due to the renormalization of $R_{\mu
\nu }R^{\mu \nu }$ and $R^{2}$ \textit{via} metric-tensor field
redefinitions, are independent of the generalization of (\ref{moller}) to
quantum field theory, so they exist also in the Schwinger-Keldysh approach.

Since causality is anyway violated in the end, it is meaningful to study a
generalization of (\ref{moller}) that is closer to the standard formulation
of quantum field theory, where correlation functions are out-in expectation
values. The prescrition adopted in this paper is to replace the right-hand
side of (\ref{moller}) with the real part of the out-in expectation value of
the stress tensor.

\bigskip

Specifically, in a fully quantized theory the quantum action $S_{q}[\varphi
_{q}]$, depending on the quantum fields $\varphi _{q}$, is defined as the
real part of the generating functional $\Gamma [\Phi ]$ of one-particle
irreducible Green functions, under the assumption that $i$) $\varphi
_{q}\equiv \Phi $ is real, if the fields $\varphi $ are real bosonic, or $ii$%
) $\varphi _{q}=\Phi $ is the conjugate of $\overline{\Phi }$, if the fields 
$\varphi $ are complex bosonic or fermionic. The variation of $S_{q}$ with
respect to $\varphi _{q}$ gives the quantum field equations. This minimum
principle applies both to fully quantized theories and to partially
classical, partially quantum theories.

Consider a quantum field theory of fields $\varphi $. I\ first assume that
the $\varphi $'s are real bosonic and later generalize the argument to the
other types of fields. Define, as usual, the generating functionals 
\[
Z[J]=\int \mathcal{D}\varphi \ \exp \left[ i\int \mathrm{d}^{4}x\ \left( 
\mathcal{L}[\varphi (x)]+J(x)\varphi (x)\right) \right] 
\]
and 
\[
W[J]=i\ln Z[J],\qquad \Gamma [\Phi ]=-W[J[\Phi ]]-\int \mathrm{d}^{4}x\
J[\Phi ](x)\ \Phi (x), 
\]
of disconnected, connected and one-particle irreducible correlation
functions, respectively, where 
\[
\Phi [J](x)=\left\langle \varphi (x)\right\rangle _{J}=-\frac{\delta W[J]}{%
\delta J(x)},\qquad J[\Phi ](x)=-\frac{\delta \Gamma }{\delta \Phi (x)}. 
\]
For the moment it is convenient to work with real fields $\varphi $. Then it
is natural to take real sources $J$.\ Nevertheless, $Z[J]$, $W[J]$ and $\Phi
[J]$ are complex functionals of $J$. The imaginary parts of the T-ordered
correlation functions are originated by the $i\varepsilon $-prescription in
the propagators. Consequently, if $J$ is real, $\Phi $ cannot be a good
quantum field and $\Gamma ,W,Z$ cannot be good quantum actions.

In a more general framework, assume that the sources $J$ are complex.
Observe that now $J$ are complex sources for real fields $\varphi $. The
T-anti-ordered Green functions are encoded in the conjugate functionals 
\begin{equation}
W^{*}[J^{*}],\quad \Phi ^{*}[J^{*}]=-\frac{\delta W^{*}[J^{*}]}{\delta J^{*}}%
,\quad \Gamma ^{*}[\Phi ^{*}]=-W^{*}[J^{*}]-\ J^{*}\cdot \Phi ^{*}.
\label{antiord}
\end{equation}
For convenience, integrals such as $\int \mathrm{d}^{4}x\ J(x)\Phi (x)$ are
often shortened as $J\cdot \Phi $.

Write $J=J_{q}+iJ_{q}^{\prime }$, where $J_{q}$, $J_{q}^{\prime }$ are real.
Now I prove that there exists a unique functional $J_{q}^{\prime }[J_{q}],$
in perturbation theory, such that $\Phi [J]$ is real.

The reality of $\Phi [J]$ is expressed by the condition 
\begin{equation}
\Phi [J]=\Phi ^{*}[J^{*}],\qquad \text{i.e. }\frac{\delta W}{\delta J}[%
J_{q}+iJ_{q}^{\prime }]=\frac{\delta W^{*}}{\delta J^{*}}[%
J_{q}-iJ_{q}^{\prime }].  \label{equ}
\end{equation}
Formula (\ref{equ}) is an equation for $J_{q}^{\prime }[J_{q}]$. Since at
the tree level $\Gamma [\Phi ]$ is real and coincides with the classical
action, $J_{q}^{\prime }$ is at least one loop. In the perturbative
expansion (\ref{equ}) reads 
\begin{equation}
J_{q}^{\prime }\left\{ \frac{\delta ^{2}W}{\delta J^{2}}[J_{q}]+\frac{\delta
^{2}W^{*}}{\delta J^{*2}}[J_{q}]\right\} =-i\Phi [J_{q}]+i\Phi ^{*}[J_{q}]+%
\mathcal{O}(J_{q}^{\prime 2}).  \label{eq1}
\end{equation}
This equation admits one solution, since $J_{q}^{\prime }$, on the left-hand
side, is multiplied by the real part of the two-point function, which is
certainly invertible. For example, in momentum space for scalar fields 
\[
\frac{\delta ^{2}W}{\delta \widetilde{J}(-p)\delta \widetilde{J}(p)}=\frac{1%
}{p^{2}-m^{2}+i\varepsilon }+\mathcal{O}(\lambda ), 
\]
where $\lambda $ collectively denotes the coupling constants that
parametrize the interactions of the theory. The left-hand side of (\ref{eq1}%
) is just 
\[
2\ \mathrm{P}\left( \frac{1}{p^{2}-m^{2}}\right) \widetilde{J}_{q}^{\prime
}(p)+\mathcal{O}(\lambda J_{q}^{\prime }), 
\]
where P denotes the principal part. Returning to coordinate space, the
solution reads 
\[
J_{q}^{\prime }[J_{q}]=-(\partial ^{2}+m^{2})\mathop{\rm Im}\Phi [J_{q}]+%
\mathcal{O}(\lambda J_{q}^{\prime },J_{q}^{\prime 2}). 
\]
The higher orders can be worked out recursively in powers of $\lambda $ and
in the loop expansion.

\bigskip

Because of its reality, the functional $\Phi [J_{q}+iJ_{q}^{\prime }[J_{q}]]$
can be taken as the quantum field $\varphi _{q}[J_{q}]$, with source $J_{q}$%
. Then the quantum action is 
\[
S_{q}[\varphi _{q}]\equiv \mathop{\rm Re}\Gamma [\varphi _{q}] 
\]
and coincides with the Legendre transform of $\mathop{\rm Re}W$, written as
a functional of $J_{q}$. Indeed, consider 
\[
W_{q}[J_{q}]\equiv \mathop{\rm Re}W[J_{q}+iJ_{q}^{\prime }[J_{q}]]. 
\]
It is immediate to show, using (\ref{equ}), that 
\[
-\frac{\delta W_{q}[J_{q}]}{\delta J_{q}}=\Phi [J_{q}+iJ_{q}^{\prime
}[J_{q}]]=\Phi ^{*}[J_{q}-iJ_{q}^{\prime }[J_{q}]]=\varphi _{q}[J_{q}]. 
\]
Then, if $J_{q}[\varphi _{q}]$ denotes the inverse of $\varphi _{q}[J_{q}]$,
the Legendre transform gives 
\[
-W_{q}[J_{q}[\varphi _{q}]]-J_{q}[\varphi _{q}]\cdot \varphi _{q}=%
\mathop{\rm Re}\Gamma [\Phi ]=S_{q}[\varphi _{q}], 
\]
as desired.

Summarizing, there exists a unique complex source $J$ such that the
functional $\Phi [J]$ is real. The quantum field $\varphi _{q}$ coincides
with $\Phi $ and the quantum action $S_{q}[\varphi _{q}]$ is just the real
part of $\Gamma [\Phi ]$.

\bigskip

The generating functional $\Gamma [\Phi ]$ can be reconstructed from the
quantum action $S_{q}[\varphi _{q}]$. Indeed, $S_{q}[\varphi _{q}]$ contains
the reals parts of the T-ordered Green functions. The imaginary parts of the
Green functions can be perturbatively calculated from the real parts.

For example, if the theory is unitary, the unitarity equation reads 
\begin{equation}
\mathop{\rm Im}T=\frac{1}{2}TT^{\dagger },  \label{unit}
\end{equation}
where $S=1+iT$ is the S-matrix, $SS^{\dagger }=1$. Since $T$ is at least of
order one in the interactions, (\ref{unit}) implies that $\mathop{\rm Im}T$
is at least of order two. So, the equation (\ref{unit}) recursively
determines the imaginary parts of the correlation functions from the
lower-order real parts.

If the theory is not unitary, a more general version of the identity (\ref
{unit}), with the same structure as (\ref{unit}), follows from the largest
time equation \cite{thooftveltman}. It cannot be interpreted as a unitarity
equation (the summation over intermediate states is affected by minus signs,
due to propagating ghosts), but it can be used to calculate the imaginary
parts of correlation functions from the lower-order real parts.

Thus, in complete generality the quantum action $S_{q}[\varphi _{q}]$
contains the full information about the theory.

\bigskip

The arguments of this section can be applied also to a partially classical,
partially quantum field theory. In that case, let $\varphi _{c}$ denote the
classical fields, with action $S_{c}[\varphi _{c}]$, and $\varphi $ the
quantized fields, with classical action $S[\varphi ,\varphi _{c}]$, embedded
in the external $\varphi _{c}$-background. The procedure described above
defines the quantum action $S_{q}[\varphi _{q},\varphi _{c}]=\mathop{\rm Re}%
\Gamma [\Phi ,\varphi _{c}]$, with $\varphi _{q}=\Phi =\mathop{\rm real}$.
The total action $S_{\text{tot}}[\varphi _{c},\varphi _{q}]$ of the
partially classical, partially quantum theory is obtained adding the
classical action $S_{c}$ of the fields $\varphi _{c}$ to $S_{q}$, namely 
\[
S_{\text{tot}}[\varphi _{c},\varphi _{q}]=S_{c}[\varphi _{c}]+S_{q}[\varphi
_{q},\varphi _{c}]. 
\]
For example, for classical gravity coupled with quantum matter, $\varphi
_{c} $ is the metric tensor $g_{\mu \nu }$, $S_{c}$ is the Einstein action,
and $S_{q}$ is the real part of the $\Gamma $ functional in external
gravity, so 
\begin{equation}
S_{\text{tot}}[g,\varphi _{q}]=\frac{1}{2\kappa ^{2}}\int \mathrm{d}^{4}x%
\sqrt{-g}\left[ R(g)-2\Lambda \right] +\mathop{\rm Re}\Gamma [\varphi
_{q},g].  \label{stot}
\end{equation}
The field equations of gravity are $\delta S_{\text{tot}}[g,\varphi
_{q}]/\delta g^{\mu \nu }=0$, namely 
\begin{equation}
R_{\mu \nu }-\frac{1}{2}g_{\mu \nu }R+g_{\mu \nu }\Lambda =-\kappa ^{2}%
\mathop{\rm Re}\left\langle T_{\mu \nu }\right\rangle ,\qquad \left\langle
T_{\mu \nu }\right\rangle =\frac{2}{\sqrt{-g}}\frac{\delta \Gamma [\varphi
_{q},g]}{\delta g^{\mu \nu }}.  \label{eqgrav}
\end{equation}
The matter field equations are $\delta S_{\text{tot}}[g,\varphi _{q}]/\delta
\varphi _{q}=0$ and have to be solved consistently with (\ref{eqgrav}). The
simplest solution is $\varphi _{q}=0$ or $\varphi _{q}=$constant (if there
is a vacuum expectation value). Then the Einstein equations (\ref{eqgrav})
describe how the spacetime geometry is affected by quantized matter fields
circulating in the loops. Together with (\ref{eqgrav}), they generalize the
M\o ller-Rosenfeld approach (\ref{moller}) to quantum field theory.

\bigskip

Working with complex bosonic fields and/or fermionic fields $\varphi $, $%
\overline{\varphi }$, denote the associated sources with $J$, $\overline{J}$%
. The functionals are $Z[J,\overline{J}]$, $W[J,\overline{J}]$ and $\Gamma
[\Phi ,\overline{\Phi }]$, which is the Legendre transform of $W[J,\overline{%
J}]$. Repeating the argument outlined above, if the source $\overline{J}$ is
the conjugate of $J$, then the functional $\overline{\Phi }[J,\overline{J}]$
is not the conjugate of $\Phi [J,\overline{J}]$. Instead, if the sources $J$%
, $\overline{J}$ are not the conjugates of each other, the relation between $%
J$ and $\overline{J}$ can be determined imposing that the functional $%
\overline{\Phi }[J,\overline{J}]$ is the conjugate of $\Phi [J,\overline{J}]$%
. In that case, the quantum action is 
\[
S_{q}[\varphi _{q},\overline{\varphi }_{q}]=\frac{1}{2}\left( \Gamma [\Phi ,%
\overline{\Phi }]+\Gamma ^{\dagger }[\overline{\Phi },\Phi ]\right) , 
\]
where $\varphi _{q}=\Phi $, $\overline{\varphi }_{q}=\overline{\Phi }$. The
other arguments extend straightforwardly.

In the presence of non-Abelian gauge fields, the gauge transformation of $%
\Phi $ can be a complex, non-local functional $\left\langle s\Phi
\right\rangle $, where $s$ denotes the BRST operator. Then the definition of
a gauge invariant real quantum functional $S_{q}[\varphi _{q}]$ is not
evident, at least in the most general framework. It is preferable to define
the functional $\Gamma [\Phi ]$ using the background field method \cite{back}%
, where now $\Phi $ denotes the background field, and the quantum field is
set to zero. The background field method ensures manifest gauge invariance
and, most of all, the gauge transformation of $\Phi $ preserves the reality
of $\Phi $. Then it is straightforward to identify $\Phi $ with the quantum
field $\varphi _{q}$ and define the quantum action $S_{q}[\varphi _{q}]$ as
the real part of $\Gamma $.

Even using the background field method, however, the functional $\Gamma $
depends on the gauge-fixing parameters. Call $G$ the unbroken non-Abelian
gauge group of the theory and $\varphi _{G}$ the fields that transform
non-trivially under $G$. The $\varphi _{G}$-quantum field equations can be
solved setting all $\varphi _{G}$'s to zero. The physical justification is
that the $G$-interactions are short-range (and even confining in QCD), so
the boundary conditions for the $\varphi _{G}$'s are that they tend to zero
with an appropriate velocity at infinity, which implies $\varphi _{G}\equiv
0 $ by the unicity of the solution.

Setting all $\varphi _{G}$'s to zero removes also the gauge-fixing
dependence of $\Gamma $. Indeed, at $\varphi _{G}=0$, the functional $\Gamma 
$ depends only on $g_{\mu \nu }$ and the other $G$-invariant fields, namely
it is a collection of correlation functions containing only insertions of $G$%
-invariant operators, so, by the usual BRST arguments, it cannot depend on
the gauge-fixing parameters. Abelian gauge fields $A$ need not be set to
zero, since $\Gamma $ is both gauge invariant and gauge-fixing independent
at $A\neq 0$.

In Euclidean theories, which are employed, for example, in the study of
critical phenomena, the average field $\Phi =\left\langle \varphi
\right\rangle _{J,\overline{J}}$ is the conjugate of $\overline{\Phi }%
=\left\langle \overline{\varphi }\right\rangle _{J,\overline{J}}$ and the
generating functionals $W[J,\overline{J}]$ and $\Gamma [\Phi ,\overline{\Phi 
}]$ are hermitian, if the sources $\overline{J}$ are the conjugates of $J$.
Then the functional $\Gamma [\Phi ,\overline{\Phi }]$ is the good quantum
action, $\Phi $ and $\overline{\Phi }$ being the quantum fields.

\section{Field redefinitions that reabsorb terms quadratically proportional
to the field equations}

\setcounter{equation}{0}

In this section I prove that a term quadratically proportional to the field
equations can be reabsorbed with a field redefinition. This theorem is used
to construct the map $\mathcal{M}$ that relates the higher-derivative theory
with the acausal theory.

Consider an action $S$ depending on the fields $\phi _{i}$, where the index $%
i$ labels both the field type, the component and the spacetime point. Add a
term quadratically proportional to the field equations $S_{i}\equiv \delta
S/\delta \phi _{i}$ and define the modified action 
\[
S^{\prime }[\phi _{i}]=S[\phi _{i}]+S_{i}F_{ij}S_{j}, 
\]
where $F_{ij}$ is symmetric and can contain derivative operators. Summation
over repeated indices (including the integration over spacetime points) is
understood. The theorem states that there exists a field redefinition 
\begin{equation}
\phi _{i}^{\prime }=\phi _{i}+\Delta _{ij}S_{j},  \label{redef}
\end{equation}
with $\Delta _{ij}$ symmetric, such that, perturbatively in $F$ and to all
orders in powers of $F$, 
\begin{equation}
S^{\prime }[\phi _{i}]=S[\phi _{i}^{\prime }].  \label{equa}
\end{equation}
Here is the proof. The condition (\ref{equa}) can be written as 
\[
S[\phi _{i}]+S_{i}F_{ij}S_{j}=S[\phi _{i}+S_{j}\Delta _{ij}]=S[\phi
_{i}]+\sum_{n=1}^{\infty }\frac{1}{n!}S_{k_{1}\cdots
k_{n}}\prod_{l=1}^{n}(\Delta _{k_{l}m_{l}}S_{m_{l}}), 
\]
after a Taylor expansion, where $S_{k_{1}\cdots k_{n}}\equiv \delta
^{n}S/(\delta \phi _{k_{1}}\cdots \delta \phi _{k_{n}})$. This equality is
verified if 
\begin{equation}
\Delta _{ij}=F_{ij}-\Delta _{k_{1}i}\Delta _{k_{2}j}\sum_{n=2}^{\infty }%
\frac{1}{n!}S_{k_{1}k_{2}k_{3}\cdots k_{n}}\prod_{l=3}^{n}(\Delta
_{k_{l}m_{l}}S_{m_{l}}),  \label{genfor}
\end{equation}
where the product is meant to be equal to unity when $n=2$. Equation (\ref
{genfor}) can be solved recursively for $\Delta $ in powers of $F$. The
first terms of the solution are 
\begin{equation}
\Delta _{ij}=F_{ij}-\frac{1}{2}F_{k_{1}i}F_{k_{2}j}S_{k_{1}k_{2}}+\cdots
\label{32}
\end{equation}

\bigskip

The theorem just proved is very general. It works both for local and
non-local theories. Assume that the spacetime dimension $d$ is greater than
two, so that the fields $\varphi $ have positive dimensionalities $%
d_{\varphi }$ in units of mass. Call ``perturbatively local'' a functional
that can be expanded in powers of the fields and their derivatives. That
means, for example, that it does not contain low-energy singularities, such
as $1/\partial _{\mu }$, $1/\Box $, etc. Call ``perturbatively local
expansion'' the expansion in powers of the fields and their derivatives. If $%
S^{\prime }[\phi _{i}]$ and $S[\phi _{i}]$ are perturbatively local, then $%
F_{xy}$ has the form 
\begin{equation}
F_{xy}=\left( f_{x}+f_{x}^{\mu }\partial _{\mu }+f_{x}^{\mu \nu }\partial
_{\mu }\partial _{\nu }+\cdots \right) \delta (x-y),  \label{Fxy}
\end{equation}
where $f_{x}^{\mu _{1}\cdots \mu _{n}}$ are perturbatively local tensorial
functionals of the fields $\phi $ and their derivatives in $x$. Now I prove
that the field redefinition (\ref{redef}) is perturbatively local, and the
solution of (\ref{genfor}) can be worked out recursively and has the same
form as (\ref{Fxy}), namely 
\begin{equation}
\Delta _{xy}=\left( g_{x}+g_{x}^{\mu }\partial _{\mu }+g_{x}^{\mu \nu
}\partial _{\mu }\partial _{\nu }+\cdots \right) \delta (x-y),\qquad
g_{x}^{\mu _{1}\cdots \mu _{k}}=f_{x}^{\mu _{1}\cdots \mu _{k}}+\mathcal{O}%
(f^{2}).  \label{Dxy}
\end{equation}

The functionals $g_{x}^{\mu _{1}\cdots \mu _{m}}$, $f_{x}^{\mu _{1}\cdots
\mu _{m}}$ have dimensionalities $2d_{\varphi }-d-m<0$. Equation (\ref
{genfor}) splits into separate equations for $g_{x}^{\mu _{1}\cdots \mu
_{k}} $, that can be solved recursively in powers of $f_{x}^{\mu _{1}\cdots
\mu _{m}}$. Each functional $f_{x}^{\mu _{1}\cdots \mu _{m}}$ can be
considered of the same order. At each order in $f$ the solution is worked
out term-by-term in the perturbatively local expansion.

Write the perturbatively local expansions of $f_{x}^{\mu _{1}\cdots \mu
_{m}} $ and $g_{x}^{\mu _{1}\cdots \mu _{m}}$ as 
\[
f_{x}^{\mu _{1}\cdots \mu _{m}}=\sum c_{f}^{(m,p,q)}\mathcal{O}_{p,q}^{\mu
_{1}\cdots \mu _{m}}[\varphi (x)]\text{,\qquad }g_{x}^{\mu _{1}\cdots \mu
_{m}}=\sum c_{g}^{(m,p,q)}\mathcal{O}_{p,q}^{\mu _{1}\cdots \mu
_{m}}[\varphi (x)], 
\]
where $\mathcal{O}_{p,q}^{\mu _{1}\cdots \mu _{m}}[\varphi ]$ denotes a
basis of local operators constructed with $p$ derivatives and $q$ fields and 
$c_{f}^{(m,p,q)}$, $c_{g}^{(m,p,q)}$ are numerical coefficients, with
dimensionalities $2d_{\varphi }-d-m-p-qd_{\varphi }<0$. Finitely many
parameters $M$ with positive dimensionalities are contained in the action $S$%
. The dimensionalities of $M$ are obviously bounded by $d$. Each term in the
sum of (\ref{genfor}) is polynomial in $M$, so (\ref{genfor}) can be
translated into equations for the $c_{g}$'s that have schematically the form 
\begin{equation}
c_{g}=c_{f}+\sum_{n=2}^{\infty }P_{n-1}(M)c_{g}^{n}\text{,}  \label{gc}
\end{equation}
where $P_{n-1}(M)$ is a polynomial of degree $n-1$ in $M$. Thus each $c_{g}$
receives $\mathcal{O}(f^{n})$ contributions from a finite number of
coefficients $c_{f}$'s, which proves that the equations (\ref{gc}) can be
solved recursively.

\bigskip

If both $S^{\prime }[\phi _{i}]$ and $S[\phi _{i}]$ are local, $F_{xy}$ is
local. Even then, in general, $\Delta _{xy}$ is only perturbatively local.
Actually, the resummation of derivatives in (\ref{Dxy}) can produce a
non-local field redefinition. Take an ordinary free field theory $S[\phi
_{i}]$. Then $S_{k_{1}\cdots k_{n}}=0$ for every $n>2$, while $%
S_{k_{1}k_{2}} $ is field-independent and quadratic in the derivatives. The
modified action $S^{\prime }[\phi _{i}]$ describes a higher-derivative
theory. Equation (\ref{genfor}) reads 
\[
\Delta _{ij}=F_{ij}-\frac{1}{2}\Delta _{k_{1}i}\Delta
_{k_{2}j}S_{k_{1}k_{2}} 
\]
and is solved in matrix form by 
\[
\Delta =\left( \sqrt{1+2FS}-1\right) S^{-1}. 
\]
Clearly, the solution $\Delta _{ij}$ is non-local, but perturbatively local.
In the next subsection these facts are illustrated explicitly for gravity in
the quadratic approximation.

\bigskip

A known situation where the theorem applies is the three-dimensional $U(1)$
gauge theory. The field equations of the Chern-Simons action 
\[
S[A]=\frac{1}{2\alpha _{\mathrm{CS}}}\int \varepsilon ^{\mu \nu \rho }F_{\mu
\nu }A_{\rho } 
\]
are $F^{\mu \nu }=0$, so there exists a field redefinition $A_{\mu }^{\prime
}(A,\alpha /\alpha _{\mathrm{CS}})$ such that 
\begin{equation}
S^{\prime }[A]=S[A^{\prime }],  \label{abi}
\end{equation}
where $S^{\prime }$ is the sum of the Chern-Simons action plus the square of
the field strength, 
\[
S^{\prime }[A]=\frac{1}{\alpha _{\mathrm{CS}}}\int \varepsilon ^{\mu \nu
\rho }F_{\mu \nu }A_{\rho }-\frac{1}{4\alpha }\int F_{\mu \nu }F^{\mu \nu }. 
\]

\subsection{The map $\mathcal{M}$ for gravity}

In pure gravity, the theorem just proved ensures that there exists a field
redefinition that maps a class of higher-derivative theories into the
Einstein theory. For example, there exists a field redefinition $%
g\rightarrow g^{\prime }(g,a,b)$ such that 
\begin{equation}
S_{\text{HD}}[g]=S_{\text{E}}[g^{\prime }],  \label{cabo}
\end{equation}
where 
\begin{eqnarray}
S_{\text{HD}}[g] &=&\frac{1}{2\kappa ^{2}}\int \sqrt{-g}\left[ R(g)+aR_{\mu
\nu }R^{\mu \nu }(g)+bR^{2}(g)\right] ,  \label{shd} \\
S_{\text{E}}[g] &=&\frac{1}{2\kappa ^{2}}\int \sqrt{-g}R(g)  \label{se}
\end{eqnarray}
Indeed, the terms $R_{\mu \nu }R^{\mu \nu }$ and $R^{2}$ are quadratically
proportional to the field equations of the action $S_{\text{E}}[g]$. The
lowest-order contributions to the map $\mathcal{M}$ are, from (\ref{redef}),
(\ref{32}), 
\begin{equation}
g_{\mu \nu }^{\prime }(g,a,b)=g_{\mu \nu }-aR_{\mu \nu }+\frac{1}{2}%
(a+2b)g_{\mu \nu }R+\mathcal{O}(a^{2},b^{2},ab).  \label{lowest}
\end{equation}

I stress once again that the identity (\ref{cabo}) does not imply that
higher-derivative gravity is physically equivalent to Einstein gravity.
Indeed, it is evident, already in the free-field limit, that the degrees of
freedom of $S_{\text{HD}}$ and $S_{\text{E}}$ are different. Nevertheless,
formula (\ref{cabo}) and the more general identity (\ref{equa}) are useful
to relate the renormalization properties of the two theories. In the next
section the identity (\ref{cabo}) is used to prove that classical gravity
coupled with quantum matter is predictive, namely all divergences are
renormalized redefining the fields and a finite number of independent
couplings.

The field redefinition $g^{\prime }(g,a,b)$ is the map $\mathcal{M}$ for
gravity. It is clearly nonlocal. When a source term is added, the map is in
general acausal (see the subsection 3.3 and section 5). Thus, in general the
map $g^{\prime }(g)$ relates higher-derivative gravity with acausal gravity.

\bigskip

In the presence of a cosmological constant, the theorem ensures that there
exists a field redefinition $g^{\prime }(g)$ such that 
\begin{equation}
S_{\text{HD}}^{(\Lambda )}[g]=S_{\text{E}}^{(\Lambda )}[g^{\prime }],
\label{accabo}
\end{equation}
where 
\begin{eqnarray*}
S_{\text{HD}}^{(\Lambda )}[g] &=&S_{\text{HD}}[g]-\frac{\Lambda }{\kappa ^{2}%
}\int \sqrt{-g}=\frac{1}{2\widetilde{\kappa }^{2}}\int \sqrt{-g}\left[
R(g)-2\Lambda +\widetilde{a}\widehat{R}_{\mu \nu }\widehat{R}^{\mu \nu }(g)+%
\widetilde{b}\widehat{R}^{2}(g)\right] , \\
S_{\text{E}}^{(\Lambda )}[g] &=&\frac{1}{2\widetilde{\kappa }^{2}}\int \sqrt{%
-g}\left[ R(g)-2\Lambda \right] ,
\end{eqnarray*}
and $\kappa ^{2}=\widetilde{\kappa }^{2}(1+2a\Lambda +8b\Lambda )$, $%
\widetilde{a}=a\widetilde{\kappa }^{2}/\kappa ^{2}$, $\widetilde{b}=b%
\widetilde{\kappa }^{2}/\kappa ^{2}$. Indeed, the hatted tensors 
\begin{equation}
\widehat{R}_{\mu \nu }=R_{\mu \nu }-g_{\mu \nu }\Lambda ,\qquad \widehat{R}%
=R-4\Lambda ,  \label{hattedR}
\end{equation}
vanish on the solutions to the field equations of $S_{\text{E}}^{(\Lambda
)}[g]$.

In the next subsection the map $g^{\prime }(g)$ is worked out explicitly in
the quadratic approximation in the absence of a cosmological constant.

\subsection{The map $\mathcal{M}$ for gravity in the quadratic approximation}

It is instructive to work out the field redefinition explicitly for gravity
in the quadratic approximation. The expansion around flat space is defined
as 
\[
g_{\mu \nu }=\eta _{\mu \nu }+2\kappa \phi _{\mu \nu }, 
\]
where $\eta _{\mu \nu }=$diag$(1,-1,-1,-1)$. The trace of $\phi _{\mu \nu }$
is denoted with $\phi $. Below, I use the convention $t_{\mu _{1}\cdots \mu
_{n}}^{2}\equiv t_{\mu _{1}\cdots \mu _{n}}t^{\mu _{1}\cdots \mu _{n}}$,
where $t_{\mu _{1}\cdots \mu _{n}}$ is any tensor.

The identity (\ref{cabo}) reads, in the quadratic approximation, 
\begin{equation}
S^{\prime }[\phi ]=S[\phi ^{\prime }],  \label{cabo2}
\end{equation}
where 
\begin{eqnarray*}
S[\phi ] &=&\frac{1}{2}\int \mathrm{d}^{4}x\left\{ (\partial _{\mu }\phi
_{\rho \sigma })^{2}-(\partial _{\mu }\phi )^{2}+2(\partial ^{\mu }\phi
)(\partial ^{\nu }\phi _{\mu \nu })-2(\partial ^{\mu }\phi _{\mu \nu
})^{2}\right\} , \\
S^{\prime }[\phi ] &=&S[\phi ]+\frac{1}{2}\int \mathrm{d}^{4}x\left\{
a\left( \Box \phi _{\mu \nu }+\partial _{\mu }\partial _{\nu }\phi -\partial
_{\mu }\partial ^{\alpha }\phi _{\nu \alpha }-\partial _{\nu }\partial
^{\alpha }\phi _{\mu \alpha }\right) ^{2}\right. \left. +4b\left( \Box \phi
-\partial ^{\mu }\partial ^{\nu }\phi _{\mu \nu }\right) ^{2}\right\} ,
\end{eqnarray*}
and the field transformation is 
\begin{equation}
\phi _{\mu \nu }=\frac{1}{\sqrt{1-a\Box }}\left( \phi _{\mu \nu }^{\prime }-%
\frac{1}{3}\eta _{\mu \nu }\phi ^{\prime }+\eta _{\mu \nu }\frac{1}{3\Box }%
\partial ^{\rho }\partial ^{\sigma }\phi _{\rho \sigma }^{\prime }\right) +%
\frac{\eta _{\mu \nu }}{3\sqrt{1-b^{\prime }\Box }}\left( \phi ^{\prime }-%
\frac{1}{\Box }\partial ^{\rho }\partial ^{\sigma }\phi _{\rho \sigma
}^{\prime }\right) ,  \label{put}
\end{equation}
where $b^{\prime }\equiv -2(a+3b)$.

It is immediate to check that 
\begin{equation}
\widetilde{\phi }_{\mu \nu }=\frac{1}{\sqrt{1-a\Box }}\widetilde{\phi }_{\mu
\nu }^{\prime },  \label{trace}
\end{equation}
where $\widetilde{\phi }_{\mu \nu }$ and $\widetilde{\phi }_{\mu \nu
}^{\prime }$ are the traceless parts of $\phi _{\mu \nu }$ and $\phi _{\mu
\nu }^{\prime }$, respectively. If $b^{\prime }=a$ the transformation (\ref
{put}) becomes simply 
\begin{equation}
\phi _{\mu \nu }=\frac{1}{\sqrt{1-a\Box }}\phi _{\mu \nu }^{\prime }.
\label{ab'}
\end{equation}

\bigskip

Due to (\ref{trace}), the gauge-fixing 
\begin{equation}
\partial ^{\mu }\widetilde{\phi }_{\mu \nu }^{\prime }=0  \label{g1}
\end{equation}
implies also 
\begin{equation}
\partial ^{\mu }\widetilde{\phi }_{\mu \nu }=0.  \label{g2}
\end{equation}
Using (\ref{g1}) and (\ref{g2}), the identity (\ref{cabo2}) simplifies to 
\[
\frac{1}{2}\int \mathrm{d}^{4}x\left\{ (\partial _{\mu }\widetilde{\phi }%
_{\rho \sigma })^{2}+a(\Box \widetilde{\phi }_{\mu \nu })^{2}-\frac{3}{8}%
\left[ (\partial _{\mu }\phi )^{2}+b^{\prime }(\Box \phi )^{2}\right]
\right\} =\frac{1}{2}\int \mathrm{d}^{4}x\left\{ (\partial _{\mu }\widetilde{%
\phi }_{\rho \sigma }^{\prime })^{2}-\frac{3}{8}(\partial _{\mu }\phi
^{\prime })^{2}\right\} 
\]
and the field redefinition (\ref{put}) becomes 
\begin{equation}
\widetilde{\phi }_{\mu \nu }=\frac{1}{\sqrt{1-a\Box }}\widetilde{\phi }_{\mu
\nu }^{\prime },\qquad \phi =\frac{1}{\sqrt{1-b^{\prime }\Box }}\phi
^{\prime }.  \label{trasfa}
\end{equation}

\subsection{Physical effects}

A\ non-renormalizable theory contains infinitely many vertices, with an
arbitrarily high number of derivatives. The usual low-energy expansion is
obtained expanding the action in powers of the fields and their momenta and
considering the (bosonic) fields and momenta of the same order. However,
sometimes it is useful to study different expansions. For example, there are
situations where it is possible to resum the expansion in powers of the
fields exactly, but it not straightforward to resum the expansion in powers
of the momenta \cite{renscal}. Here, instead, the expansion in powers of the
fields is difficult to resum, but it is straightforward to resum certain
expansions in powers of the momenta, which lead for example to the square
roots of formulas (\ref{put}) and (\ref{tensa}). The resummation of momenta
is meaningful in a regime in which the fields are weak, but not necessarily
slowly varying, where it is sufficient to keep only the linear and quadratic
terms in $\phi ^{\prime }$.

Thus, to illustrate the effects on interactions in the weak-field
approximation, add a source term 
\begin{equation}
S_{\text{source}}[\phi ,T]=-\kappa \int \mathrm{d}^{4}x~\phi _{\mu \nu
}T^{\mu \nu },  \label{s1}
\end{equation}
where $T_{\mu \nu }$ is the energy-momentum tensor. Then (\ref{cabo2})
extends to 
\begin{equation}
S_{\mathrm{HD}}[\phi ,T]=S_{\mathrm{AC}}[\phi ^{\prime },T],  \label{cabo3}
\end{equation}
where 
\[
S_{\mathrm{HD}}[\phi ,T]=S^{\prime }[\phi ]+S_{\text{source}}[\phi
,T],\qquad S_{\mathrm{AC}}[\phi ^{\prime },T]=S[\phi ^{\prime }]+S_{\text{%
source}}[\phi ^{\prime },T^{\prime }(T)] 
\]
and 
\begin{equation}
T_{\mu \nu }^{\prime }(T)=\frac{1}{\sqrt{1-a\Box }}\left( T_{\mu \nu }-\frac{%
1}{3}\eta _{\mu \nu }T+\frac{1}{3\Box }\partial _{\mu }\partial _{\nu
}T\right) +\frac{1}{3\sqrt{1-b^{\prime }\Box }}\left( \eta _{\mu \nu }T-%
\frac{1}{\Box }\partial _{\mu }\partial _{\nu }T\right) ,  \label{tensa}
\end{equation}
$T$ being the trace of $T_{\mu \nu }$. The expansions of (\ref{put}) and (%
\ref{tensa}) in powers of $a$ and $b^{\prime }$ are perturbatively local, in
agreement with the conclusions derived previously. At the non-perturbative
level in $a$ and $b^{\prime }$, the operators 
\begin{equation}
\frac{1}{\sqrt{1-a\Box }},\qquad \frac{1}{\sqrt{1-b^{\prime }\Box }}
\label{ope}
\end{equation}
stand for convolutions with the generalized functions 
\begin{equation}
\mathcal{C}_{4}^{(f)}(x)=\int \frac{\mathrm{d}^{4}k}{(2\pi )^{4}}\frac{%
\mathrm{e}^{-ikx}}{\sqrt{1+fk^{2}}},  \label{attra}
\end{equation}
where $f=$ $a,b^{\prime }$. The operator $1/\Box $ in (\ref{tensa}) stands
for the convolution with 
\begin{equation}
\mathcal{G}_{4}(t,\mathbf{x})=\frac{1}{4\pi |\mathbf{x}|}\delta (t-|\mathbf{x%
}|).  \label{g4}
\end{equation}
The Fourier transforms (\ref{attra}) need prescriptions for the contour
integrations. The prescriptions must ensure that the $f\rightarrow 0$ limits
of $\mathcal{C}_{4}^{(f)}(x)$ are regular, for the reasons explained below.

The action $S_{\mathrm{AC}}[\phi ^{\prime },T]$ couples the field $\phi
_{\mu \nu }^{\prime }$ with $T_{\mu \nu }^{\prime }(T)$, which is a sort of
spacetime average of the matter stress tensor $T_{\mu \nu }$. In section 5
it is shown that if $f$ is negative $\mathcal{C}_{4}^{(f)}$ admits a real
causal prescription. That prescription, however, does not survive the
radiative corrections and ultimately the value of $T_{\mu \nu }^{\prime }(t,%
\mathbf{x})$ at time $t$ depend also on the spacetime points that are
located in the future light cone of $x$ or are spacelike separated from $x$.
Then, causality is violated.

If a complex prescription is used for (\ref{attra}), the conclusions of the
previous section apply, and the tree-level quantum action $S_{q\mathrm{AC}%
}[\phi ^{\prime },T]$ is the real part of $S_{\mathrm{AC}}[\phi ^{\prime
},T] $, with the convention that the quantum field $\varphi _{q}\equiv \phi
^{\prime }$ is real. Neither the choice of the prescription, nor the
suppression of the imaginary part of $S_{\mathrm{AC}}[\phi ^{\prime },T]$,
affect the perturbative expansion in powers of $a$ and $b^{\prime }$ and the
renormalizability of the theory, discussed in the next section.

Note that resummations similar to the ones that lead to (\ref{ope}) are
familiar in high-energy physics, where they are produced by the
renormalization group. Specifically, the renormalization group is able to
resum certain expansions in powers of the couplings and the logarithms of
momenta. In gravity the coupling is, in some sense, itself a momentum. Then
the gravitational analogue is the resummation of an expansion in powers of
momenta and the logarithms of momenta. In section 5 the radiative
corrections are included, and produce the expected dependence on the
logarithms of momenta, see formulas (\ref{trasfanuova}) and (\ref{cc}).

\bigskip

The identity (\ref{cabo3}) is the map $\mathcal{M}$ for classical gravity in
the weak-field approximation. The action $S_{\mathrm{HD}}$ contains
higher-derivative kinetic terms, while the action $S_{\mathrm{AC}}$ does
not. Now, assume that the physical theory is $S_{q\mathrm{AC}}[\phi ^{\prime
},T]$. That means that the spacetime geometry is described by $\phi _{\mu
\nu }^{\prime }$ and the source of the physical interaction is $T_{\mu \nu }$%
. However, the spacetime geometry is not affected directly by $T_{\mu \nu }$%
. Instead, it is sensitive to the ``effective stress-tensor'' $\func{Re}%
T_{\mu \nu }^{\prime }$, which is a spacetime average of $T_{\mu \nu }$.
Observe that $\func{Re}T_{\mu \nu }^{\prime }$ need not obey the positivity
constraints obeyed by $T_{\mu \nu }$. Using the gauge-fixing 
\begin{equation}
\partial ^{\nu }\phi _{\mu \nu }^{\prime }=\frac{1}{2}\partial _{\mu }\phi
^{\prime },  \label{gfix}
\end{equation}
the gravitational field equations read 
\begin{equation}
\Box \phi _{\mu \nu }^{\prime }=-\kappa \func{Re}T_{\mu \nu }^{\prime }(T)+%
\frac{\kappa }{2}\eta _{\mu \nu }\func{Re}T^{\prime }(T).  \label{box}
\end{equation}
Equation (\ref{box}) is a second-order partial differential equation and
must be supplemented with the usual boundary conditions, e.g. $\phi _{\mu
\nu }^{\prime }(t_{0},\mathbf{x})$ and $\partial _{0}\phi _{\mu \nu
}^{\prime }(t_{0},\mathbf{x})$ at the initial time $t_{0}$.

It is instructive to compare equation (\ref{box}) with the equation
generated by the higher-derivative theory. Assume that the physical theory
is $S_{\mathrm{HD}}[\phi ,T]$. Then, with the gauge-fixing analogous to (\ref
{gfix}), the field equation for $\phi _{\mu \nu }$ reads 
\begin{equation}
\Box \phi _{\mu \nu }-\frac{1}{2}\eta _{\mu \nu }\Box \phi -a\left( \Box
^{2}\phi _{\mu \nu }+\frac{1}{2}\eta _{\mu \nu }\Box ^{2}\phi -\Box \partial
_{\mu }\partial _{\nu }\phi \right) -2b\left( \Box \eta _{\mu \nu }-\partial
_{\mu }\partial _{\nu }\right) \Box \phi =-\kappa T_{\mu \nu }.  \label{hdeq}
\end{equation}
This equation is a fourth-order partial differential equation and must be
supplemented with unusual boundary conditions, e.g. $\partial _{0}^{n}\phi
_{\mu \nu }^{\prime }(t_{0},\mathbf{x})$, $n=0,1,2,3$ at the initial time $%
t_{0}$. It has extra solutions that (\ref{box}) does not have. In
particular, $\Box \phi _{\mu \nu }\neq 0$ even at $T_{\mu \nu }=0$. Thus,
equations (\ref{box}) and (\ref{hdeq}) are physically inequivalent.

When the higher-derivative local equation (\ref{hdeq}) is converted into the
second-order non-local equation (\ref{box}) by the map $\mathcal{M}$, the
extra solutions of (\ref{hdeq}) disappear. They are killed by the
requirement that the generalized functions (\ref{attra}) be regular in the
limit $f\rightarrow 0$. In practice, the map $\mathcal{M}$ consists of a
universal choice of the extra boundary conditions, which suppresses the
unwanted degrees of freedom, but in general produces causality violations.
These ideas are inspired by known treatments of the Abraham-Lorentz force in
classical electrodynamics \cite{jackson}, which are reviewed in section 5.
To be precise, a certain ambiguity survives also in (\ref{box}), due to the
freedom to choose different prescriptions for $\mathcal{C}_{4}^{(f)}(x)$.

\bigskip

The causality violations can be physically tested studying, for example, the
gravitational force predicted by $S_{q\mathrm{AC}}[\phi ^{\prime },T]$.
Consider a set of small rigid spheres of masses $m_{i}$ and radii $R_{i}$,
moving along trajectories $\mathbf{r}_{i}(t)$. The mass distributions are $%
\rho _{i}(\mathbf{x}-\mathbf{r}_{i})$, where $\rho _{i}(\mathbf{r}%
)=3m_{i}/(4\pi R_{i}^{3})$ for $|\mathbf{r}|\leq R_{i}$ and $\rho _{i}(%
\mathbf{r})=0$ for $|\mathbf{r}|>R_{i}$. The stress-energy tensor reads 
\begin{equation}
T^{\mu \nu }(t,\mathbf{x})=\sum_{i}\frac{\rho _{i}(\mathbf{x}-\mathbf{r}%
_{i}(t))}{\sqrt{1-\mathbf{\dot{r}}_{i}^{2}(t)}}v_{i}^{\mu }(t)v_{i}^{\nu
}(t),  \label{tmunu}
\end{equation}
where $v_{i}^{\mu }(t)=(1,\mathbf{\dot{r}}_{i}(t))$. The total action,
including the kinetic terms of the spheres, is 
\begin{equation}
S_{\text{tot}}[\phi ^{\prime },\mathbf{r}_{i}]=S_{q\mathrm{AC}}[\phi
^{\prime },T]-\sum_{i}m_{i}\int \mathrm{d}t\sqrt{1-\mathbf{\dot{r}}%
_{i}^{2}(t)}.  \label{action}
\end{equation}
The equations of motions of $S_{\text{tot}}[\phi ^{\prime },\mathbf{r}_{i}]$
are involved, but some qualitative aspects of their solutions can be studied
in the non-relativistic limit, where the time derivatives of (\ref{box}) and
(\ref{hdeq}) are negligible and the causality violations disappear. The
stress tensor (\ref{tmunu}) simplifies to 
\[
T_{00}(\mathbf{x})=\sum_{i}m_{i}\rho _{i}(\mathbf{x}-\mathbf{r}_{i}), 
\]
any other component being negligible. From (\ref{nonrela}), the generalized
functions (\ref{attra}) become 
\begin{equation}
\mathcal{C}_{4}^{(f)}(x)\rightarrow -\frac{\delta (t)K_{1}(r/\sqrt{-f})}{%
2\pi ^{2}fr}.  \label{nonrela2}
\end{equation}
For concreteness, assume that the spheres are pointlike, $\rho _{i}(\mathbf{r%
})=m_{i}\delta ^{(3)}(\mathbf{r})$. Using (\ref{nonrela2}), $T_{\mu \nu
}^{\prime }$ has components 
\[
T_{00}^{\prime }(\mathbf{x})=\sum_{i}\frac{2}{3}\overline{\rho }_{i}^{(a)}(%
\mathbf{x})+\frac{1}{3}\overline{\rho }_{i}^{(b^{\prime })}(\mathbf{x}%
),\qquad \overline{\rho }_{i}^{(f)}(\mathbf{x})=\frac{m_{i}K_{1}(|\mathbf{x}-%
\mathbf{r}_{i}|/\sqrt{-f})}{2\pi ^{2}(-f)|\mathbf{x}-\mathbf{r}_{i}|}, 
\]
and 
\[
T_{ij}^{\prime }(\mathbf{x})=\frac{1}{3}\left( \delta _{ij}-\frac{\partial
_{i}\partial _{j}}{\triangle }\right) \left( \overline{\rho }_{i}^{(a)}(%
\mathbf{x})-\overline{\rho }_{i}^{(b^{\prime })}(\mathbf{x})\right) , 
\]
while $T_{i0}^{\prime }=0$. In practice, a pointlike sphere effectively
smears out into distributions of mass $\overline{\rho }_{i}^{(f)}(\mathbf{x}%
) $, which are sensibly different from zero in regions of radii $\sqrt{|f|}$.

The force is $\mathbf{F}_{i}=-\mathbf{\nabla }_{i}U$. The potential energy $%
U $ can be read from 
\begin{equation}
\frac{1}{2}\func{Re}S_{\text{source}}[\phi ^{\prime },T^{\prime }]=-\int 
\mathrm{d}t\,U,  \label{potential}
\end{equation}
and $\phi _{\mu \nu }^{\prime }$ can be calculated from (\ref{box}), using (%
\ref{g4}). The factor one half in (\ref{potential}) is because $\phi _{\mu
\nu }^{\prime }$ is proportional to $T_{\mu \nu }^{\prime }$. The result is 
\begin{equation}
U=-\frac{\kappa ^{2}}{8\pi }\int ~\frac{\mathrm{d}^{3}\mathbf{x\,}\mathrm{d}%
^{3}\mathbf{x}^{\prime }}{|\mathbf{x-x}^{\prime }|}\left( \func{Re}T_{\mu
\nu }^{\prime }(\mathbf{x})\func{Re}T^{\prime \,\mu \nu }(\mathbf{x}^{\prime
})-\frac{1}{2}\func{Re}T^{\prime }(\mathbf{x})\func{Re}T^{\prime }(\mathbf{x}%
^{\prime })\right) \mathrm{.}  \label{U}
\end{equation}
For $a,b^{\prime }<0$ the integral gives 
\begin{equation}
U=-\frac{\kappa ^{2}}{8\pi }\sum_{i<j}\frac{m_{i}m_{j}}{r_{ij}}\left( 1-%
\frac{4}{3}\mathrm{e}^{-r_{ij}/\sqrt{-a}}+\frac{1}{3}\mathrm{e}^{-r_{ij}/%
\sqrt{-b^{\prime }}}\right) ,\qquad r_{ij}=|\mathbf{r}_{i}-\mathbf{r}_{j}|,
\label{staticplot}
\end{equation}
where the self-energies have been subtracted away.

The generalization of $U$ when either $a$ or $b^{\prime }$ is positive is
simple, but left to the reader.

So far, the Newton law has been verified down to about 0.1 millimeters \cite
{gravitytests} without observing any deviations, so the experimental bound
on the values of $|a|$ and $|b^{\prime }|$ is 
\begin{equation}
|a|,|b^{\prime }|<2.5\cdot 10^{5}(\mathrm{eV})^{-2}.  \label{bondi}
\end{equation}

\section{Renormalization of classical gravity coupled with quantum matter}

\setcounter{equation}{0}

In this section I show that the divergences of acausal gravity coupled with
quantum matter can be removed with a finite number of independent couplings
without introducing higher-derivative terms in the gravitational sector. The
map $\mathcal{M}$ is used to relate the renormalization of acausal gravity
coupled with quantum matter to the renormalization of higher-derivative
gravity coupled with quantum matter.

For simplicity, I first consider a theory that does not contain parameters
with positive dimensionality in units of mass. The generalization to
theories with cosmological constant, masses and super-renormalizable
couplings is described later on. Moreover, I use the dimensional
regularization technique, which is BRST invariant and does not produce
power-like divergences.

The classical action is written as 
\begin{equation}
S_{\text{AC}}[g,\varphi ,\lambda ,\lambda ^{\prime },\kappa ]=\frac{1}{%
2\kappa ^{2}}\int \mathrm{d}^{4}x\sqrt{-g}R+S_{m}[\varphi ,g,\lambda
]+\Delta S_{m}[\varphi ,g,\lambda ,\lambda ^{\prime }].  \label{acca2}
\end{equation}
Here $S_{m}$ collects the power-counting renormalizable terms of the matter
action embedded in external gravity and $\lambda $ denotes the dimensionless
couplings of $S_{m}$. For example, in the case of (massless) QED $S_{m}$ is
equal to 
\begin{equation}
S_{m}=\int \mathrm{d}^{4}x\sqrt{-g}\left( -\frac{1}{4}F_{\mu \nu }F^{\mu \nu
}+\overline{\psi }iD\mathcal{\!\!\!\!}\slash \psi \right) ,  \label{qe}
\end{equation}
where $D_{\mu }=\mathcal{D}_{\mu }+ieA_{\mu }$ is the covariant derivative.
In the case of scalar fields, the action $S_{m}$ includes also the
non-minimal term $R\varphi ^{2}$: 
\begin{equation}
S_{m}=\int \mathrm{d}^{4}x\sqrt{-g}\left( \frac{1}{2}g^{\mu \nu }(\partial
_{\mu }\varphi )(\partial _{\nu }\varphi )-\frac{1+2\eta }{12}R\varphi ^{2}-%
\frac{\lambda }{4!}\varphi ^{4}\right) .  \label{scal}
\end{equation}

In (\ref{acca2}) $\Delta S_{m}$ collects the terms of dimensionality greater
than four, parametrized by couplings $\lambda ^{\prime }$ with negative
dimensionalities in units of mass.

In four dimensions, neither $S_{m}$ nor $\Delta S_{m}$ include pure-gravity
terms, namely $S_{m}[0,g,\lambda ]=\Delta S_{m}[0,g,\lambda ,\lambda
^{\prime }]=0$. In higher dimensions this requirement has to be
appropriately modified (see below).

The theory is renormalizable if the correction $\Delta S_{m}$ to the matter
action is such that the divergences of (\ref{acca2}) are subtracted away
renormalizing the couplings of (\ref{acca2}) and redefining the fields. The
field redefinition of $g_{\mu \nu }$ cannot depend on the matter fields,
because the matter fields are quantized (they are integrated in the
functional integral), while the metric tensors $g_{\mu \nu }$ is just an
external source.

\subsection{Renormalizability of the higher-derivative theory}

Before proving the renormalizability of the acausal theory, I\ recall the
properties of the higher-derivative theory. The action of the
higher-derivative classical gravity coupled with quantum matter is

\begin{equation}
S_{\text{HD}}[\overline{g},\varphi ,\lambda ,a,b,\kappa ]=\frac{1}{2\kappa
^{2}}\int \mathrm{d}^{4}x\sqrt{-\overline{g}}\left( \overline{R}+a\overline{R%
}_{\mu \nu }\overline{R}^{\mu \nu }+b\overline{R}^{2}\right) +S_{m}[\varphi ,%
\overline{g},\lambda ],  \label{hg}
\end{equation}
where $\overline{R}_{\mu \nu }=R_{\mu \nu }(\overline{g})$. The metric
tensor is denoted with $\overline{g}$ to distinguish it from the metric
tensor $g$ of the theory (\ref{acca2}).

As in the theory (\ref{acca2}), only the matter fields $\varphi $ are
quantized. By power-counting, the renormalization of $S_{m}$ generates
counterterms of dimensionality four, which can be grouped in four classes:
counterterms proportional to the terms of $S_{m}$, counterterms proportional
to the field equations, BRST-exact counterterms and pure-gravity
counterterms. The pure-gravity counterterms are 
\begin{equation}
\int \mathrm{d}^{4}x\sqrt{-\overline{g}}\left( \alpha \overline{R}_{\mu \nu
\rho \sigma }\overline{R}^{\mu \nu \rho \sigma }+\beta \overline{R}_{\mu \nu
}\overline{R}^{\mu \nu }+\gamma \overline{R}^{2}\right) ,  \label{gb}
\end{equation}
but, as usual, the first term of this list is converted into a combination
of the other two, up to a total derivative, using the Gauss-Bonnet identity.
Thus, the higher-derivative theory (\ref{hg}) can be renormalized redefining
the matter fields $\varphi $ and the parameters $\lambda $, $a$ and $b$. The 
$\varphi $-redefinition is just multiplicative ($\varphi _{\mathrm{B}%
}=Z_{\varphi }^{1/2}\varphi $), so it does not depend on the gravitational
background.

The bare action reads 
\begin{equation}
S_{\text{HD~}\mathrm{B}}=S_{\text{HD}}[\overline{g},\varphi _{\mathrm{B}%
},\lambda _{\mathrm{B}},a_{\mathrm{B}},b_{\mathrm{B}},\kappa ]=\frac{1}{%
2\kappa ^{2}}\int \mathrm{d}^{4}x\sqrt{-\overline{g}}\left( \overline{R}+a_{%
\mathrm{B}}\overline{R}_{\mu \nu }\overline{R}^{\mu \nu }+b_{\mathrm{B}}%
\overline{R}^{2}\right) +S_{m}[\varphi _{\mathrm{B}},\overline{g},\lambda _{%
\mathrm{B}}].  \label{sb}
\end{equation}
There is no need to redefine the metric tensor and the Newton constant, so $%
\overline{g}_{\mu \nu \mathrm{B}}=\overline{g}_{\mu \nu }$ and $\kappa _{%
\mathrm{B}}=\kappa $.

Finally, the generating functional $\Gamma _{\text{HD}}[\overline{g},\Phi
,\lambda ,a,b,\kappa ]$ of one-particle irreducible Green functions is
defined by 
\[
\int \mathcal{D}\varphi \ \mathrm{\exp }\left( iS_{\text{HD~}\mathrm{B}%
}+i\int \sqrt{-\overline{g}}J\varphi \right) =\exp \left( i\Gamma _{\text{HD}%
}[\overline{g},\Phi ,\lambda ,a,b,\kappa ]+i\int \sqrt{-g}J\Phi \right) , 
\]
where $J=-(1/\sqrt{-g})(\delta \Gamma _{\text{HD}}/\delta \Phi )$.

\subsection{Usage of the map $\mathcal{M}$}

The next step is to use the map $\mathcal{M}$ (\ref{cabo}) to convert the
higher-derivative theory (\ref{hg}) into a theory of the form (\ref{acca2}).
Call $\overline{G}(g,a,b)$ the function such that for $\overline{g}=%
\overline{G}(g,a,b)$%
\begin{equation}
\int \mathrm{d}^{4}x\sqrt{-\overline{g}}\left[ \overline{R}(\overline{g})+a%
\overline{R}_{\mu \nu }\overline{R}^{\mu \nu }(\overline{g})+b\overline{R}%
^{2}(\overline{g})\right] =\int \mathrm{d}^{4}x\sqrt{-g}R(g),  \label{ride2}
\end{equation}
where the bar on the curvature tensors means that they are those of the
metric $\overline{g}$, and define the correction $\Delta S_{m}$ as 
\begin{equation}
\Delta S_{m}[\varphi ,g,\lambda ,a,b]=S_{m}[\varphi ,\overline{G}%
(g,a,b),\lambda ]-S_{m}[\varphi ,g,\lambda ].  \label{ride3}
\end{equation}
Then, (\ref{ride2}) and (\ref{ride3})\ ensure that 
\begin{equation}
S_{\text{HD}}[\overline{G}(g,a,b),\varphi ,\lambda ,a,b,\kappa ]=S_{\text{AC}%
}[g,\varphi ,\lambda ,a,b,\kappa ],  \label{ide}
\end{equation}
where the parameters $\lambda ^{\prime }$ of (\ref{acca2}) are just $a$ and $%
b$, whose dimensionalities are $-2$.

The acausal theory $S_{\text{AC}}[g,\varphi ,\lambda ,a,b,\kappa ]$ does not
contain higher-derivatives terms in the pure-gravity sector. However, due to 
$\Delta S_{m}$, the matter sector contains vertices that depend
non-polynomially on the gravitational field and its derivatives.

The main properties of $\Delta S_{m}$ can be read directly from (\ref{ide}).
First, the vertices of $\Delta S_{m}$ have dimensionality greater than four.
They are constructed with the matter fields, the Ricci tensor and their
covariant derivatives. Moreover, they are proportional to the Ricci tensor
and polynomial in the matter fields (and their covariant derivatives), of
the same degree as $S_{m}$. Clearly, $\Delta S_{m}[0,g,\lambda ,a,b]=0$.

Using (\ref{lowest}) the lowest-order contributions to the correction $%
\Delta S_{m}$ are 
\begin{eqnarray}
\Delta S_{m} &=&\Delta S_{m}^{\text{(HEAD)}}+\Delta S_{m}^{\text{(QUEUE)}}, 
\nonumber \\
\Delta S_{m}^{\text{(HEAD)}} &=&\int \mathrm{d}^{4}x\sqrt{-g}\ \left[ -\frac{%
a}{2}T_{m}^{\mu \nu }R_{\mu \nu }+\frac{1}{4}(a+2b)RT_{m}\right] ,
\label{head} \\
\Delta S_{m}^{\text{(QUEUE)}} &=&\mathcal{O}(a^{2},b^{2},ab),  \nonumber
\end{eqnarray}
where $T_{m}^{\mu \nu }=-(2/\sqrt{-g})(\delta S_{m}/\delta g_{\mu \nu })$ is
the stress-tensor of the uncorrected matter sector and $T_{m}$ denotes its
trace. Formula (\ref{head}) clarifies the meaning of the couplings $a$ and $%
b $ in the acausal theory: they multiply the vertices that couple $%
T_{m}^{\mu \nu }$ to the Ricci tensor. The other contributions to $\Delta
S_{m}$ are either proportional to $T_{m}^{\mu \nu }$ times derivatives of
the Ricci tensor or quadratically proportional to the Ricci tensor.

The correction $\Delta S_{m}$ falls in the class of non-renormalizable
perturbations constructed in ref.s \cite{renscal,nonre}. In (\ref{head}),
the terms $\Delta S_{m}^{\text{(HEAD)}}$ have dimensionality $6$. They are
multiplied by independent couplings, $a$ and $b$, and form the \textit{head}
of the perturbation. The terms $\Delta S_{m}^{\text{(HEAD)}}$ have
dimensionalities greater than $6$ and form the \textit{queue} of the
perturbation. Although the queue contains infinitely many vertices, it
contains only a finite number of independent matter operators, generated by
the functional derivatives of $T_{m}^{\mu \nu }$ with respect to the metric.
The queue does not contain new independent couplings. Its vertices are
multiplied by functions of the other couplings ($a$, $b$, $\lambda $ and $%
\kappa $), determined by certain RG\ consistency conditions, called \textit{%
reduction equations}, ensuring that the divergences of the theory are
removed renormalizing the couplings $a$, $b$, $\lambda $ and $\kappa $,
together with field redefinitions.

\subsection{Renormalizability}

The renormalizability of (\ref{acca2}) is proved using the renormalizability
of the higher-derivative theory (\ref{hg}) and the map $\mathcal{M}$.
Briefly, the divergences of $S_{\text{HD}}[\overline{g},\varphi ,\lambda
,a,b,\kappa ]$ are renormalized redefining $\varphi $, $\lambda $, $a$ and $%
b $ at fixed $\overline{g}_{\mu \nu }$ and $\kappa $: since $g$ is a
function of $\overline{g}$, $a$ and $b$, the divergences of $S_{\text{AC}%
}[g,\varphi ,\lambda ,a,b,\kappa ]$ are removed redefining $g$, $\varphi $, $%
\lambda $, $a$ and $b$ at fixed $\kappa $.

The acausal theory is renormalizable if there exists a bare metric tensor $%
g_{\mathrm{B}}$, depending only on $g$ and the couplings, such that the bare
action 
\begin{equation}
S_{\text{AC~}\mathrm{B}}\equiv S_{\text{AC}}[g_{\mathrm{B}},\varphi _{%
\mathrm{B}},\lambda _{\mathrm{B}},a_{\mathrm{B}},b_{\mathrm{B}},\kappa ]
\label{barac}
\end{equation}
produces finite Green functions. The $g$-redefinition $g_{\mathrm{B}}$ that
does this job is obtained solving the condition 
\begin{equation}
\overline{G}(g_{\mathrm{B}},a_{\mathrm{B}},b_{\mathrm{B}})=\overline{G}%
(g,a,b).  \label{fieldredef}
\end{equation}
More explicitly, calling $g=G(\overline{g},a,b)$ the inverse of $\overline{g}%
=\overline{G}(g,a,b)$, 
\begin{equation}
g_{\mathrm{B}}=G(\overline{G}(g,a,b),a_{\mathrm{B}},b_{\mathrm{B}})=g+%
\mathcal{O}(\hbar ),  \label{riden}
\end{equation}
and, to the lowest order, using (\ref{lowest}), 
\[
g_{\mu \nu \mathrm{B}}=g_{\mu \nu }+(a-a_{\mathrm{B}})R_{\mu \nu }+\frac{1}{2%
}(a_{\mathrm{B}}-a+2b_{\mathrm{B}}-2b)g_{\mu \nu }R+\hbar \mathcal{O}(\hbar
,a,b). 
\]
Observe that the couplings $a$ and $b$ cancel out in the lowest-order
expression, which confirms that $g_{\mu \nu \mathrm{B}}$ is truly a field
redefinition, not a redefinition of the couplings.

Using (\ref{barac}), (\ref{ide}), (\ref{fieldredef}) and (\ref{sb}), $S_{%
\text{AC~}\mathrm{B}}$ is equal to 
\begin{equation}
S_{\text{AC~}\mathrm{B}}=S_{\text{HD}}[\overline{G}(g_{\mathrm{B}},a_{%
\mathrm{B}},b_{\mathrm{B}}),\varphi _{\mathrm{B}},\lambda _{\mathrm{B}},a_{%
\mathrm{B}},b_{\mathrm{B}},\kappa ]=S_{\text{HD}}[\overline{G}%
(g,a,b),\varphi _{\mathrm{B}},\lambda _{\mathrm{B}},a_{\mathrm{B}},b_{%
\mathrm{B}},\kappa ]=S_{\text{HD\ }\mathrm{B}}.  \label{two}
\end{equation}
This equality ensures that the set of Feynman diagrams of the acausal theory
is obtained from the set of diagrams of the higher-derivative theory, once $%
\overline{g}$ on the external legs is replaced with the finite function $%
\overline{G}(g,a,b)$. Thus, the Green functions of the acausal theory are
finite and collected in the generating functional 
\begin{equation}
\Gamma _{\text{AC}}[g,\Phi ,\lambda ,a,b,\kappa ]=\Gamma _{\text{HD}}[%
\overline{G}(g,a,b),\Phi ,\lambda ,a,b,\kappa ].  \label{gamma}
\end{equation}

According to the arguments of section 2, the quantum action $S_{q\text{AC}%
}[g,\varphi _{q},\lambda ,a,b,\kappa ]$ is the real part of $\Gamma _{\text{%
AC}}[g,\Phi ,\lambda ,a,b,\kappa ]$, with the convention that $i$) $g$ is
real and $ii$) $\Phi =$ $\varphi _{q}$ is real if the fields $\varphi $ are
real bosonic, while $\Phi $ is the conjugate of $\overline{\Phi }$ if the
fields $\varphi $, $\overline{\varphi }$ are complex bosonic or fermionic.
The gravitational field equations are given by the variation of $S_{q\text{AC%
}}[g,\varphi _{q},\lambda ,a,b,\kappa ]$ with respect to the metric tensor $%
g_{\mu \nu }$. The variation of $S_{q\text{AC}}[g,\varphi _{q},\lambda
,a,b,\kappa ]$ with respect to $\varphi _{q}$ generates the quantum field
equations of matter.

\bigskip

Thus the theory $S_{q\text{AC}}[g,\varphi _{q},\lambda ,a,b,\kappa ]$ is a
predictive formulation of classical gravity coupled with quantum matter. No
higher-derivative kinetic terms have been added to the pure-gravity sector.
The number of independent couplings is finite: the gravitational couplings
are just three, namely the Newton constant $\kappa $, which does not
renormalize, plus $a$ and $b$; on the other hand, the number of couplings $%
\lambda $ belonging to the matter sector is constrained by power counting.

\bigskip

In the presence of a cosmological constant, the identity (\ref{accabo}) has
to be used. If the matter sector does not contain parameters with positive
dimensionalities in units of mass, the parameters $a$, $b$ renormalize
exactly as above, and $\kappa $, $\Lambda $ do not renormalize. In the
acausal theory, the true Newton constant is $\widetilde{\kappa }$, which
gets renormalized because it is a function of $a$, $b$, $\kappa $ and $%
\Lambda $. Observe that the cosmological constant $\Lambda $ is the same on
the two sides of the map $\mathcal{M}$. If the matter sector contains
parameters with positive dimensionalities in units of mass, then there are
independent renormalizations of the Newton constant and the cosmological
constant.

\subsection{Classical gravity coupled with quantum matter in higher
dimensions}

The construction of this section can be generalized to higher dimensions.
Assume first that the cosmological constant is zero and the matter sector
does not contain parameters of positive dimensionalities in units of mass.

On the higher-derivative side of the map $\mathcal{M}$, the counterterms can
be classified in two subsets, in connection with the expansion of the metric
tensor around flat space, $\overline{g}_{\mu \nu }=\eta _{\mu \nu }+h_{\mu
\nu }$ : $i$) ``kinetic counterterms'', namely counterterms that contain
contributions quadratic in $h$; \noindent $ii$) ``vertex counterterms'',
namely counterterms that do not contain contributions quadratic in $h$.

The kinetic counterterms can always be converted into counterterms
quadratically proportional to the Einstein field equations \cite{abse},
which can be reabsorbed by the map $\mathcal{M}$, plus vertex counterterms.
The vertex counterterms can be quadratically proportional to the Einstein
field equations or not. Those that are can be reabsorbed by the map $%
\mathcal{M}$, those that are not must be included in $S_{m}$, multiplied by
independent couplings. With these arrangements the theorem of section 2
applies.

For example, in six dimensions \cite{vannieu}, kinetic counterterms of
dimensionality 6 are 
\begin{equation}
\int \sqrt{-\overline{g}}\ \overline{R}_{\mu \nu \rho \sigma }\overline{%
\bigtriangledown }_{\alpha }\overline{\bigtriangledown }^{\alpha }\overline{R%
}^{\mu \nu \rho \sigma },\qquad \int \sqrt{-\overline{g}}\ \overline{%
\bigtriangledown }_{\alpha }\overline{R}_{\mu \nu \rho \sigma }\overline{%
\bigtriangledown }^{\mu }\overline{R}^{\alpha \nu \rho \sigma },  \label{ty1}
\end{equation}
etc. Using partial integrations and Bianchi identities, and commuting
derivatives, these counterterms can be converted into terms quadratically
proportional to the Ricci tensor, which can be reabsorbed by the map $%
\mathcal{M}$, plus vertex counterterms. Vertex counterterms are

\begin{equation}
\int \sqrt{-\overline{g}}\ \overline{R}_{\mu \nu }^{\quad \rho \sigma }%
\overline{R}_{\alpha \beta }^{\quad \mu \nu }\overline{R}_{\rho \sigma
}^{\quad \alpha \beta },\qquad \int \sqrt{-\overline{g}}\ \overline{R}_{\mu
}^{\nu }\overline{R}_{\nu \alpha }^{\quad \rho \sigma }\overline{R}_{\rho
\sigma }^{\quad \mu \alpha },\qquad \int \sqrt{-\overline{g}}\ \overline{R}%
_{\mu }^{\nu }\overline{R}_{\rho }^{\mu }\overline{R}_{\nu }^{\rho },
\label{ty2}
\end{equation}
etc. The third of (\ref{ty2}) can be reabsorbed by the map $\mathcal{M}$,
while the other two must be included in $S_{m}$, multiplied by independent
couplings.

The vertex counterterms can be ignored in the quadratic analysis of
causality violations (see the next section).

\bigskip

In the presence of a cosmological constant, or if the matter sector contains
parameters with positive dimensionalities in units of mass (which generate
the cosmological constant by renormalization), it is convenient to expand
the metric $\overline{g}_{\mu \nu }=$ $g_{\mu \nu }^{(0)}+h_{\mu \nu }$
around a maximally symmetric metric $g_{\mu \nu }^{(0)}$, such that 
\[
R_{\mu \nu \rho \sigma }^{(0)}=\frac{2\Lambda }{(d-1)(d-2)}\left( g_{\mu
\rho }^{(0)}g_{\nu \sigma }^{(0)}-g_{\mu \sigma }^{(0)}g_{\nu \rho
}^{(0)}\right) , 
\]
where $d$ is the spacetime dimension. The gravitational counterterms are
more conveniently rearranged as functions of the hatted Riemann tensor 
\[
\widehat{\overline{R}}_{\mu \nu \rho \sigma }=\overline{R}_{\mu \nu \rho
\sigma }-\frac{2\Lambda }{(d-1)(d-2)}(\overline{g}_{\mu \rho }\overline{g}%
_{\nu \sigma }-\overline{g}_{\mu \sigma }\overline{g}_{\nu \rho }) 
\]
and its covariant derivatives, because $\widehat{\overline{R}}_{\mu \nu \rho
\sigma }$ vanishes on the metric $g_{\mu \nu }^{(0)}$. Again, the
counterterms can be distinguished into kinetic counterterms (those that
contain contributions quadratic in $h$) and vertex counterterms (those that
do not contain contributions quadratic in $h$). It was shown in \cite{abse}
that the kinetic counterterms can be converted into terms quadratically
proportional to $\widehat{\overline{R}}_{\mu \nu }$ or its covariant
derivatives, which are reabsorbed by the map $\mathcal{M}$, plus vertex
counterterms, plus a linear combination of $\overline{R}$ and $1$, that
renormalize the Newton constant and the cosmological constant. There is only
one case where this fact is not obvious, namely $\widehat{\overline{R}}_{\mu
\nu \rho \sigma }\widehat{\overline{R}}^{\mu \nu \rho \sigma }$. However,
the combination 
\begin{eqnarray*}
\widehat{\overline{\mathrm{G}}} &=&\widehat{\overline{R}}_{\mu \nu \rho
\sigma }\widehat{\overline{R}}^{\mu \nu \rho \sigma }-4\widehat{\overline{R}}%
_{\mu \nu }\widehat{\overline{R}}^{\mu \nu }+\widehat{\overline{R}}^{2}+%
\frac{8(d-3)}{(d-1)(d-2)}\Lambda \left( \overline{R}-2\Lambda \right) \\
&=&\overline{R}_{\mu \nu \rho \sigma }\overline{R}^{\mu \nu \rho \sigma }-4%
\overline{R}_{\mu \nu }\overline{R}^{\mu \nu }+\overline{R}^{2}-4\frac{%
(d-3)(d-4)}{(d-1)(d-2)}\Lambda (\overline{R}-\Lambda ),
\end{eqnarray*}
does not contain $h$-quadratic contributions, thanks a peculiar identity 
\cite{abse}, 
\[
\int \sqrt{-g}\ \widehat{\overline{\mathrm{G}}}=\frac{32(d-3)}{(d-1)(d-2)^{2}%
}\Lambda ^{2}\int \sqrt{-g^{(0)}}+\mathcal{O}\left( h^{3}\right) , 
\]
which proves that $\widehat{\overline{\mathrm{G}}}$ is a vertex counterterm.
The counterterms $\widehat{\overline{R}}_{\mu \nu \rho \sigma }\overline{%
\nabla }^{\lambda _{1}}\cdots \overline{\nabla }^{\lambda _{n}}\widehat{%
\overline{R}}_{\alpha \beta \gamma \delta },$ $\hbox{$n>0$}$, with indices
contracted in all possible ways, can be reduced by means of repeated partial
integrations, commutations of covariant derivatives and applications of the
Bianchi identities.

The counterterm $\int \sqrt{-g}\ \widehat{\overline{\mathrm{G}}}$ and the
other vertex counterterms cannot, in general, be reabsorbed by the map $%
\mathcal{M}$. They have to be included in $S_{m}$, multiplied by independent
couplings. If the matter sector is a power-counting renormalizable theory
(which, in $d>4$, means just a free theory, or the $\varphi ^{3}$ theory in
five and six dimensions), embedded in curved space, then the action $S_{%
\text{HD}}^{(\Lambda )}$ contains a finite number of terms, therefore the
acausal theory $S_{\text{AC}}^{(\Lambda )}$ has a finite number of
independent couplings.

\bigskip

Finally, in three spacetime dimensions the action (\ref{acca2}) is
renormalizable with $\Delta S_{m}=0$. Indeed, a three-dimensional
power-counting renormalizable theory in curved space generates no
higher-derivative pure-gravity counterterm: the Lorentz Chern-Simons term is
protected \cite{3dgrav}; all other higher-derivative terms constructed with
the Riemann and Ricci tensor have at least dimensionality four. So, there is
no causality violation in three dimensions. In higher dimensions
higher-derivative terms can be generated, but they must be multiplied by
parameters with positive odd dimensionalities. If such parameters are not
contained in $S_{m}$, then causality is not violated. If such parameters are
contained in $S_{m}$, then the procedure of even-dimensional theories has to
be applied and there are causality violations.

\section{Higher time derivatives, instabilities and causality violations}

\setcounter{equation}{0}

In general, the map $\mathcal{M}$ converts a causal classical theory with
instabilities, originated by higher derivatives in the kinetic term, into an
acausal classical theory without instabilities. This section is devoted to
study these properties in more detail, including the effects of the
radiative corrections.

To illustrate the logic of the discussion, it is convenient to recall the
analysis of the Abraham-Lorentz force (see for example \cite{jackson}). In
classical electrodynamics an effective description of the Larmor formula 
\begin{equation}
P=m\tau a^{2},\qquad \tau =\frac{2e^{2}}{3mc^{3}},  \label{tau}
\end{equation}
for the radiation power emitted by an accelerated particle in the adiabatic
approximation is provided by the higher-derivative equation 
\begin{equation}
ma(t)=m\tau \dot{a}(t)+F(t),  \label{AL}
\end{equation}
where $a$ is the acceleration and $F(t)$ is an external force. The term $%
m\tau \dot{a}$ is the Abraham-Lorentz force. Equation (\ref{AL}) can be
integrated one time, to give 
\begin{equation}
ma(t)=-\frac{1}{\tau }\int_{-\infty }^{t}\mathrm{d}t^{\prime }~\text{e}%
^{(t-t^{\prime })/\tau }F(t^{\prime })+ma_{0}\text{e}^{t/\tau },
\label{runa}
\end{equation}
where $a_{0}$ is the arbitrary constant. The solution (\ref{runa}) is
causal, since it depends only on the force $F(t^{\prime })$ at earlier times 
$t^{\prime }<t$. The second term is a runaway solution, which is the sign of
instability. It is present even when there are no external forces.

Observe that in (\ref{runa}) the contribution of the force at earlier times $%
t^{\prime }<t$ is exponentially amplified. The reason is that the limits $%
\tau \rightarrow 0$ of equation (\ref{AL}) and its solution (\ref{runa}) are
singular. However, physics suggests that such a limit should exist, since $%
\tau $ in (\ref{tau}) is proportional to the square of the charge.

The $\tau \rightarrow 0$ limit becomes regular only if the constant $a_{0}$
is set equal to 
\[
a_{0}=\frac{1}{m\tau }\int_{-\infty }^{\infty }\mathrm{d}t^{\prime }~\text{e}%
^{-t^{\prime }/\tau }F(t^{\prime }). 
\]
Then (\ref{runa}) becomes 
\begin{equation}
ma(t)=\frac{1}{\tau }\int_{t}^{\infty }\mathrm{d}t^{\prime }~\text{e}%
^{(t-t^{\prime })/\tau }F(t^{\prime }).  \label{AL2}
\end{equation}
The $\tau \rightarrow 0^{+}$ limit of this equation is $F=ma$, as desired.
Equation (\ref{AL2}) is a physically reasonable replacement of the
Abraham-Lorentz force. However, (\ref{AL2}) is not equivalent to (\ref{AL}).
Every solution of (\ref{AL2}) solves (\ref{AL}), but not vice versa. The
runaway solution is eliminated and at $F=0$ the acceleration vanishes. The
effective force felt by the particle is a time average of the true force $F$%
. The acceleration of the particle at a time $t$ depends on the force $F$ at
future times $t^{\prime }>t$, so causality is violated. Summarizing, the
physics described by equation (\ref{AL}) is causal but unstable, while the
physics of equation (\ref{AL2}) is stable but acausal.

The causality violations are short-range, the range being of the order $%
\Delta t\sim \tau $. Numerically, $\Delta t\sim 10^{-22}$sec. Since quantum
effects become important already at time intervals of the order of $137\tau $%
, the causality violations predicted by equation (\ref{AL2}) are
unobservable.

Writing 
\begin{equation}
ma=\frac{1}{1-\tau \frac{\mathrm{d}}{\mathrm{d}t}}F,  \label{implicit}
\end{equation}
it becomes evident that the runaway solution, which is the zero mode of $%
1-\tau \mathrm{d}/\mathrm{d}t$, is lost in the inversion of this operator,
demanding the regularity of the $\tau \rightarrow 0$ limit. The inversion of 
$1-\tau \mathrm{d}/\mathrm{d}t$ is the map $\mathcal{M}$ that relates the
equations (\ref{AL}) and (\ref{AL2}).

Observe that the presence of instabilities, or causality violations, is
related to the sign of $\tau $. No instability nor causality violation
occurs for $\tau <0$.

The rearrangement of equation (\ref{AL}) into formula (\ref{implicit}),
interpreted in the usual low-energy expansion, which throws away the
unstable solutions, is known in the literature as the \textit{regular
reduction} of the order of the differential equation and can be done also
for gravity \cite{bel,parkersimon}.

Although inspired by the arguments just recalled, the map $\mathcal{M}$
differs from the regular reduction in a crucial way. The regular reduction
is not a field redefinition, but a manipulation of the field equations. In
the case of gravity, the analogue of this operation \cite{bel} is a
manipulation of the field equations of higher-derivative gravity, coupled
with classical or quantum matter, which leaves the metric tensor unchanged.
It is not known how to implement the regular reduction for gravity at the
level of the action. The construction of this paper, instead, is performed
at the level of the action and implemented by iterative field redefinitions
of the metric tensor that renormalize the counterterms $R_{\mu \nu }R^{\mu
\nu }$ and $R^{2}$. Typical signs of the difference between the two
approaches are the square roots of (\ref{ope}), which appear naturally in
the map $\mathcal{M}$, but do not appear in (\ref{implicit}) and in the
approach of \cite{bel}. Observe that it is not possible to derive the
Abraham-Lorentz force from an action, which is why now I\ abandon this
analogy and proceed with the description of the approach of this paper in
lagrangian models.

\bigskip

Consider the higher-derivative theory 
\[
\mathcal{L}^{\prime }(q)=\frac{m}{2}\dot{q}^{2}+\frac{m\alpha ^{2}}{2}\ddot{q%
}^{2}\equiv \mathcal{L}(q)+\Delta \mathcal{L}(q),\qquad \mathcal{L}(q)=\frac{%
m}{2}\dot{q}^{2}. 
\]
The term $\Delta \mathcal{L}$ is quadratically proportional to the field
equations of $\mathcal{L}$. The map $\mathcal{M}$ is 
\begin{equation}
q(q^{\prime })=\frac{1}{\sqrt{1-\alpha ^{2}\frac{d^{2}}{dt^{2}}}}~q^{\prime
},  \label{rede}
\end{equation}
so that 
\begin{equation}
\int \mathrm{d}t\ \mathcal{L}^{\prime }(q)=\int \mathrm{d}t\ \mathcal{L}%
(q^{\prime }).  \label{id}
\end{equation}
More explicitly, 
\begin{equation}
q(t)=\int_{-\infty }^{+\infty }\mathrm{d}t^{\prime }~\mathcal{C}(t-t^{\prime
})q^{\prime }(t^{\prime }),\qquad \mathcal{C}(t)=\int_{-\infty }^{+\infty }%
\frac{\mathrm{d}k}{2\pi }\frac{e^{-ikt}}{\sqrt{1+\alpha ^{2}k^{2}}}.
\label{inte}
\end{equation}
According to the discussion about the Abraham-Lorentz force, the map $%
\mathcal{M}$ should tend to the identity in the limit $\alpha \rightarrow 0$%
, which is implicit in (\ref{inte}). Nevertheless, there might exist
different prescriptions to define $\mathcal{C}(t)$. Every prescription has
the same perturbative expansion in powers of $\alpha $.

When $\alpha ^{2}>0$ (assume $\alpha >0$ without loss of generality) the
solutions of the field equations of $\mathcal{L}^{\prime }(q)$ and $\mathcal{%
L}(q^{\prime })$ read 
\begin{equation}
q(t)=at+b+c\text{e}^{t/\alpha }+d\text{e}^{-t/\alpha },\qquad q^{\prime
}(t)=a^{\prime }t+b^{\prime },  \label{asso}
\end{equation}
respectively. At finite non-vanishing $\alpha $, $q(t)$ contains two
solutions (one of which is runaway) that are absent in $q^{\prime }(t)$.
When $\alpha ^{2}<0$ the exponentials are replaced by sine and cosine
functions and there is no runaway solution. Finally, $q(t)$ is singular in
the limit $\alpha \rightarrow 0$.

If the system is subject to an external time-dependent force $F(t)$, the
lagrangian 
\begin{equation}
\mathcal{L}^{\prime }(q,F)=\frac{m}{2}\dot{q}^{2}+\frac{m\alpha ^{2}}{2}%
\ddot{q}^{2}+qF(t),  \label{hd}
\end{equation}
is mapped by (\ref{rede}) into 
\begin{equation}
\mathcal{L}(q^{\prime },F^{\prime })=\frac{m}{2}\dot{q}^{\prime 2}+q^{\prime
}F^{\prime }(t),\qquad F^{\prime }(t)=\frac{1}{\sqrt{1-\alpha ^{2}\frac{d^{2}%
}{dt^{2}}}}F(t)=\int_{-\infty }^{+\infty }\mathrm{d}t^{\prime }~\mathcal{C}%
(t-t^{\prime })F(t^{\prime }).  \label{FF}
\end{equation}
so 
\begin{equation}
\mathcal{\qquad }\int \mathrm{d}t\mathcal{\ L}^{\prime }(q(q^{\prime
}),F)=\int \mathrm{d}t\ \mathcal{L}(q^{\prime },F^{\prime }(F)).
\label{acau}
\end{equation}

Consider the function $\mathcal{C}(t)$ in (\ref{inte}) and (\ref{FF}). For $%
\alpha ^{2}>0$ the integral (\ref{inte}) is convergent and gives 
\begin{equation}
\mathcal{C}(t)=\frac{1}{\pi |\alpha |}K_{0}\left( \frac{|t|}{|\alpha |}%
\right) ,\qquad \text{if }\alpha ^{2}>0.  \label{ar2}
\end{equation}
Causality is violated, since $F^{\prime }(t)$ depends on the force $%
F(t^{\prime })$ at future times $t^{\prime }$. The range of the causality
violations is $\Delta t=|\alpha |$.

When $\alpha ^{2}<0$ it is necessary to specify a prescription for the
contour integration in the complex plane. There is a real causal
prescription, which gives the retarded function $\mathcal{C}_{\mathrm{ret}%
}(t)$, 
\begin{equation}
\mathcal{C}_{\mathrm{ret}}(t)=\int_{-\infty }^{+\infty }\frac{\mathrm{d}k}{%
2\pi }\frac{e^{-ikt}}{\sqrt{1+\alpha ^{2}(k+i\varepsilon )^{2}}}=\frac{%
\theta (t)}{|\alpha |}J_{0}\left( \frac{|t|}{|\alpha |}\right) ,\qquad \text{%
if }\alpha ^{2}<0.  \label{ar3}
\end{equation}
The advanced function is $\mathcal{C}_{\mathrm{adv}}(t)=\mathcal{C}_{\mathrm{%
ret}}(-t)$. A complex acausal prescription is studied below in arbitrary
spacetime dimensions: see formula (\ref{cnf}).

Summarizing, the theory $\mathcal{L}(q^{\prime },F^{\prime })$ has no
unstable solution. It violates causality for $\alpha ^{2}>0$ and admits a
causal prescription for $\alpha ^{2}<0$.

Again, the redefinition (\ref{inte}) maps two physically inequivalent
theories. Once it is known whether $q$ or $q^{\prime }$ are the physical
fields, and whether $F$ or $F^{\prime }$ are the physical forces, the
physics follows from the appropriate lagrangian, (\ref{hd}) or (\ref{acau}).

\subsection{Fields}

Consider the scalar theory 
\begin{equation}
\mathcal{L}^{\prime }(\varphi ,J)=\frac{1}{2}(\partial _{\mu }\varphi
)(\partial ^{\mu }\varphi )+\frac{1}{2}\alpha ^{2}(\Box \varphi
)^{2}+\varphi J  \label{luno}
\end{equation}
in $n$ spacetime dimensions. The map 
\begin{equation}
\varphi ^{\prime }=\sqrt{1-\alpha ^{2}\Box }\varphi  \label{rede2}
\end{equation}
relates (\ref{luno}) with the theory 
\begin{equation}
\mathcal{L}(\varphi ^{\prime },J^{\prime }(J))=\frac{1}{2}(\partial _{\mu
}\varphi ^{\prime })(\partial ^{\mu }\varphi ^{\prime })+\varphi ^{\prime
}J^{\prime },\qquad J^{\prime }(x)=\int \mathrm{d}^{n}x^{\prime }~\mathcal{C}%
_{n}(x-x^{\prime })J(x^{\prime }),  \label{id2}
\end{equation}
where 
\begin{equation}
\mathcal{C}_{n}(x)=\int \frac{\mathrm{d}^{n}k}{(2\pi )^{n}}\frac{\mathrm{e}%
^{-ik\cdot x}}{\sqrt{1+\alpha ^{2}k^{2}}}.  \label{ava}
\end{equation}
Again, the regularity of the $\alpha \rightarrow 0$ limit is understood in (%
\ref{ava}).

Fields of higher spins can be treated similarly. In every case, the function 
$\mathcal{C}_{n}(x)$ is the essential ingredient of the map $\mathcal{M}$.
For gravity in the quadratic approximation the map $\mathcal{M}$ is
collected in formulas (\ref{put}) and (\ref{tensa}) and involves the
functions $\mathcal{C}_{n}(x)$ with $\alpha ^{2}$ equal to $a$ or $b^{\prime
}$.

The Fourier transform (\ref{ava}) has to be defined with an appropriate
prescription. It is convenient to begin with the prescription \cite{gelfand} 
\begin{equation}
\mathcal{C}_{n}^{\text{F}}(x)=\int \frac{\mathrm{d}^{n}k}{(2\pi )^{n}}\frac{%
\mathrm{e}^{-ik\cdot x}}{\sqrt{1+\alpha ^{2}k^{2}+i\varepsilon }}=\frac{%
\mathrm{e}^{-i\frac{\pi }{4}\left[ 2(n-1)+(n-2)(\mathrm{sign}(\alpha
^{2})-1)\right] }}{2^{(n-1)/2}\pi ^{(n+1)/2}|\alpha |^{n}}\frac{%
K_{(n-1)/2}\left( \sqrt{\frac{x^{2}}{\alpha ^{2}}-i\varepsilon ^{\prime }}%
\right) }{\left( \frac{x^{2}}{\alpha ^{2}}-i\varepsilon ^{\prime }\right)
^{(n-1)/4}},  \label{cnf}
\end{equation}
which illustrates the main features of the function $\mathcal{C}_{n}(x)$.
Observe that $\mathcal{C}_{n}^{\text{F}}(x)$ is complex, but recall that the
quantum action $S_{q\text{AC}}$ is the real part of the functional $\Gamma _{%
\text{AC}}$ of (\ref{gamma}).

In even dimensions the function $\mathcal{C}_{n}^{\text{F}}(x)$ is quite
simple. For example, 
\begin{equation}
\mathcal{C}_{4}^{\text{F}}(x)=\frac{\mathrm{sign}(\alpha ^{2})}{4\pi
^{2}|\alpha |^{4}}\frac{i\exp \left( -\sqrt{\frac{x^{2}}{\alpha ^{2}}%
-i\varepsilon }\right) }{\left( \frac{x^{2}}{\alpha ^{2}}-i\varepsilon
\right) ^{3/2}}\left( 1+\sqrt{\frac{x^{2}}{\alpha ^{2}}-i\varepsilon }%
\right) .  \label{c4}
\end{equation}
The exponential tends to zero or rapidly oscillates for $|x^{2}|\gg |\alpha
^{2}|$, so the causality violations can be experimentally tested only at
distances of the order of 
\begin{equation}
\Delta x\sim 2\pi |\alpha |  \label{range}
\end{equation}
and become physically unobservable at distances much larger than this bound.

For $\alpha ^{2}=-\overline{\alpha }^{2}<0$ there is a real causal
prescription, namely the retarded function 
\begin{equation}
\mathcal{C}_{n}^{\mathrm{ret}}(x)=\int \frac{\mathrm{d}^{n}k}{(2\pi )^{n}}%
\frac{\mathrm{e}^{-ik\cdot x}}{\sqrt{1-\overline{\alpha }^{2}\left(
k_{0}+i\varepsilon \right) ^{2}+\overline{\alpha }^{2}\mathbf{k}^{2}}},
\label{crnet}
\end{equation}
which vanishes for $t<0$. Indeed, the branch cuts are located in the lower
half $k_{0}$-plane and if $t<0$ it is possible to close the contour of
integration in the upper half plane ($\mathop{\rm Im}k^{0}>0$). By Lorentz
invariance, every point outside the light-cone admits a reference frame in
which $t<0$, so $\mathcal{C}_{n}^{\mathrm{ret}}(x)$ vanishes identically
outside the light-cone. The advanced function $\mathcal{C}_{n}^{\mathrm{adv}%
}(x)$ is defined as in (\ref{crnet}) with $k_{0}+i\varepsilon \rightarrow
k_{0}-i\varepsilon $. In the non-relativistic limit, 
\begin{equation}
\mathcal{C}_{n}(x)\rightarrow \frac{2\delta (t)K_{n/2-1}(r/\overline{\alpha }%
)}{(2\pi \overline{\alpha })^{n/2}r^{n/2-1}},  \label{nonrela}
\end{equation}
independently of the prescription, where $x=(t,\mathbf{x})$ and $r=|\mathbf{x%
}|$. For $\alpha ^{2}>0$ no causal prescription exists. When $\alpha
\rightarrow 0$ the functions $\mathcal{C}_{n}(x)$ tend to $(2\pi )^{n}\delta
^{n}(x)$, independently of the prescription. When $\alpha ^{2}\rightarrow
\infty $ they tend to zero.

\bigskip

The map $\mathcal{M}:S_{\mathrm{HD}}\rightarrow S_{\mathrm{AC}}$ is
essentially classical, because it applies to a classical theory, or to the
classical sector of a partially classical, partially quantum theory. The
generalization of the map $\mathcal{M}$ to quantum gravity should convert
higher-derivative quantum gravity into acausal quantum gravity, preserving
the renormalizability. Higher-derivative quantum gravity is renormalizable,
but not unitary \cite{stelle}. The violation of unitarity is exhibited by
the propagation of ghosts, which are the quantum counterparts of the
classical instabilities. However, the renormalization of higher-derivative
quantum gravity is singular in the limit where $a$, $b^{\prime }$ tend to
zero. It has been remarked above that the smoothness of these limits is an
essential ingredient for the map $\mathcal{M}$, to trade the instabilities
for causality violations (check the discussion about the $\tau \rightarrow 0$
limit of the Abraham-Lorentz force).

The quantum map $\mathcal{M}$ should be able convert unitarity violations
into causality violations, preserving the renormalization structure. Once
again, the map $\mathcal{M}$ cannot be a field redefinition, because a field
redefinition preserves the renormalization structure, but does not change
the poles of the S-matrix elements (see \cite{weinberg}). A naive
application of the map $\mathcal{M}$ in the functional integral restores the
ghosts by means of the Jacobian determinant. In conclusion, the construction
of a good map $\mathcal{M}$ for quantum gravity has to be left to future
investigations.

\subsection{Effects of the radiative corrections}

At the tree level, the presence of causality violations depends on the sign
of $\alpha ^{2}$. When $\alpha ^{2}<0$, causal prescriptions exist for $%
\mathcal{C}_{n}(x)$, when $\alpha ^{2}<0$ there is no causal prescription.
Beyond the tree-level, the logarithmic corrections spoil the causal
prescriptions. Nevertheless, the causality violations affect only high
energies.

Consider classical gravity coupled with a renormalizable quantum field
theory. At one loop the $a$-running is governed by the trace anomaly of the
matter sector in curved space. To the lowest order the beta functions are 
\cite{hathrell,fradkin} 
\begin{equation}
\frac{1}{\kappa ^{2}}\beta _{a}=-4c+\mathcal{O}(\lambda ),\qquad \frac{1}{%
\kappa ^{2}}\beta _{b^{\prime }}=\kappa ^{2}\mathcal{O}(\lambda ),
\label{running}
\end{equation}
where $\lambda $ denotes the matter couplings, including the parameter $\eta 
$ of formula (\ref{scal}), and 
\begin{equation}
c=\frac{12n_{v}+6n_{f}+n_{s}}{120(4\pi )^{2}},  \label{ecci}
\end{equation}
where $n_{s}$ is the number real scalars, $n_{f}$ is the number of Dirac
fermions and $n_{v}$ is the number of vector fields. For example, in QED 
\[
\beta _{a}=-\frac{3\kappa ^{2}}{5(4\pi )^{2}}-\frac{7}{9}\frac{\kappa
^{2}e^{2}}{(4\pi )^{4}}+\kappa ^{2}\mathcal{O}(e^{4}),\qquad \beta
_{b^{\prime }}=-\frac{16}{27}\frac{e^{6}\kappa ^{2}}{(4\pi )^{8}}+\kappa ^{2}%
\mathcal{O}(e^{8}). 
\]

To illustrate the effects of radiative corrections it is sufficient to
concentrate on the first contributions to the beta functions. Assume that
the interactions of the matter sector are switched off ($\lambda =0$),
namely that the matter sector is a free-field theory in curved space. Then
the renormalization of $a$ and $b^{\prime }$ is exact, 
\begin{equation}
\beta _{a}=-4c\kappa ^{2},\qquad \beta _{b^{\prime }}=0.  \label{run}
\end{equation}
The exactness of formulas (\ref{run}) holds in a larger class of models,
those whose matter sector is a conformal field theory $\mathcal{C}$ embedded
in external gravity. Then $c$ is not (\ref{ecci}), but a characteristic
quantity of $\mathcal{C}$, called ``central charge'' $c$ (see for example
ref.s \cite{centralcharges} for definitions and properties). According to (%
\ref{run}), the parameter $b^{\prime }$ does not run, but $a$ does. The $a$%
-running is 
\begin{equation}
a(-p^{2})=\overline{a}-2c\kappa ^{2}\ln \frac{-p^{2}}{\mu ^{2}}=-2c\kappa
^{2}\ln \frac{-p^{2}}{\Lambda ^{2}},\qquad \Lambda \equiv \mu \ \mathrm{\exp 
}\left( \frac{\overline{a}}{4c\kappa ^{2}}\right) ,  \label{exa}
\end{equation}
where $\overline{a}=a(\mu ^{2})$ and $\Lambda $ is the energy scale at which
the running coupling switchs its sign.

I stress again that (\ref{exa}) is an exact formula for an important class
of models. Thus, it is mandatory to investigate the physical effects of the $%
a$-running at the non-perturbative level in $a$ and $\kappa ^{2}$.

Write the higher-derivative action $S_{\text{HD}}(g)$ (\ref{shd}) as

\[
S_{\text{HD}}[g]=\frac{1}{2\kappa ^{2}}\int \sqrt{-g}\left[ R+\frac{a}{2}%
W^{\mu \nu \rho \sigma }W_{\mu \nu \rho \sigma }-\frac{b^{\prime }}{6}%
R^{2}\right] , 
\]
where $W^{\mu \nu \rho \sigma }$ is the Weyl tensor. The one-loop quantum
functional $\Gamma $ reads, in the gravity sector, 
\[
\Gamma _{\text{HD}}=\frac{1}{2\kappa ^{2}}\int \sqrt{-g}\left[ R-c\kappa
^{2}W^{\mu \nu \rho \sigma }\ln \left( \frac{\Box }{\Lambda ^{2}}\right)
W_{\mu \nu \rho \sigma }-\frac{b^{\prime }}{6}R^{2}\right] , 
\]
up to cubic terms in the curvature tensors.

In the quadratic approximation, with the gauge fixing (\ref{g1})-(\ref{g2}),
the map $\mathcal{M}$ relating the higher-derivative quantum functional $%
\Gamma _{\text{HD}}$ with the acausal functional $\Gamma _{\text{AC}}$, 
\begin{eqnarray*}
\Gamma _{\text{HD}} &=&\frac{1}{2}\int \mathrm{d}^{4}x\left\{ (\partial
_{\mu }\widetilde{\phi }_{\rho \sigma })^{2}-2c\kappa ^{2}(\Box \widetilde{%
\phi }_{\mu \nu })\ln \left( \frac{\Box }{\Lambda ^{2}}\right) (\Box 
\widetilde{\phi }^{\mu \nu })-\frac{3}{8}\left[ (\partial _{\mu }\phi
)^{2}+b^{\prime }(\Box \phi )^{2}\right] \right\} \\
&=&\frac{1}{2}\int \mathrm{d}^{4}x\left\{ (\partial _{\mu }\widetilde{\phi }%
_{\rho \sigma }^{\prime })^{2}-\frac{3}{8}(\partial _{\mu }\phi ^{\prime
})^{2}\right\} =\Gamma _{\text{AC}},
\end{eqnarray*}
is promoted, by dimensional transmutation, to the renormalization-group
invariant form 
\begin{equation}
\widetilde{\phi }_{\mu \nu }=\frac{1}{\sqrt{1+2c\kappa ^{2}\Box \ln \left( 
\frac{\Box }{\Lambda ^{2}}\right) }}\widetilde{\phi }_{\mu \nu }^{\prime
},\qquad \phi =\frac{1}{\sqrt{1-b^{\prime }\Box }}\phi ^{\prime }.
\label{trasfanuova}
\end{equation}
According to the arguments of section 2, the quantum action $S_{\text{AC}}$
is the real part of $\Gamma _{\text{AC}}$, with the convention that $\phi
^{\prime }$ is real.

The function $\mathcal{C}_{4}(x)$ mapping the traceless part $\widetilde{%
\phi }_{\mu \nu }$ is 
\begin{equation}
\mathcal{C}_{4}(x)=\int \frac{\mathrm{d}^{4}p}{(2\pi )^{4}}\frac{\mathrm{e}%
^{-ip\cdot x}}{\sqrt{1+2\xi \frac{-p^{2}}{\Lambda ^{2}}\ln \frac{-p^{2}}{%
\Lambda ^{2}}}},  \label{cc}
\end{equation}
where $\xi =$ $c\kappa ^{2}\Lambda ^{2}$. The prescription for $\ln \left(
\Box /\Lambda ^{2}\right) $ is determined by the Feynman prescription for
the propagators of the matter fields that circulate in the loops, so in
momentum space 
\[
\ln \left( \frac{-p^{2}}{\Lambda ^{2}}\right) \rightarrow \ln \left( \frac{%
-p^{2}-i\varepsilon }{\Lambda ^{2}}\right) . 
\]
Closing the contour of the $p_{0}$-integration in the upper half $p_{0}$%
-plane at infinity, the phase of $p_{0}$ ranges from $0$ to $\pi $ and the
phase of $p^{2}$ crosses the branch cut of the logarithm. The function $%
\mathcal{C}_{4}(x)$ receives a contribution from the integral along the cut
and does not vanish for $t<0$. Moreover, since the integral along the cut is
not purely imaginary, even $\func{Re}\mathcal{C}_{4}(x)$ is non-vanishing
for $t<0$. Other non-vanishing contributions can come from the cuts of the
square root. In conclusion, causality is violated due to the radiative
corrections.

A prescription for the other factor of $p^{2}$ in (\ref{cc}) can be obtained
generalizing the prescription (\ref{cnf}). Then $\mathcal{C}_{4}(x)$ is 
\begin{equation}
\mathcal{C}_{4}(x)=\int \frac{\mathrm{d}^{4}p}{(2\pi )^{4}}\frac{\mathrm{e}%
^{-ip\cdot x}}{\sqrt{1-2\xi \frac{p^{2}+i\varepsilon }{\Lambda ^{2}}\ln
\left( -\frac{p^{2}+i\varepsilon }{\Lambda ^{2}}\right) }}.  \label{cx}
\end{equation}

At sufficiently low energies, the function $\mathcal{C}_{4}(x)$ is not
sensibly different from the identity $(2\pi )^{4}\delta (x)$. The acausal
behavior can be observed starting from energies $E$ such that 
\[
E^{2}a(E^{2})\sim 1. 
\]
So far, the gravitational force has been tested down to distances of the
order of 0.1 millimeters, which means energies about $2\cdot 10^{-3}$eV,
without observing acausal behaviors. Thus the value $a(E^{2})$ of the
coupling $a$ at that energy is bounded by 
\[
|a(E)|<2.5\cdot 10^{5}\text{(eV)}^{-2}. 
\]

\section{Conclusions}

\setcounter{equation}{0}

I have proved that classical gravity coupled with quantized matter can be
renormalized with a finite number of independent couplings, without adding
higher-derivative terms to the gravitational sector. Instead, the theory
contains vertices that couple the matter stress-tensor with the Ricci tensor
and predicts the violation of causality at small distances.

The proof of renormalizability uses a map $\mathcal{M}$ that relates acausal
gravity with higher-derivative gravity. The map $\mathcal{M}$, inspired by
known treatments of the Abraham-Lorentz force in classical electrodynamics,
trades the instabilities due to higher-derivatives for causality violations.
The field equations of a partially classical, partially quantum field theory
follow from a suitable minimization principle.

The matter sector is an ordinary power-counting renormalizable theory in
curved space, with couplings $\lambda $, plus a non-renormalizable
perturbation, made by a head and a queue. The head contains two vertices
that couple the matter stress-tensor with the Ricci tensor, multiplied by
independent couplings $a$ and $b^{\prime }$. The queue contains an infinity
of higher-dimensioned vertices, polynomial in the matter fields, but
non-polynomial in the gravitational field and its derivatives. The queue
does not contain new independent couplings, rather its vertices are
multiplied by appropriate functions of the other couplings, such that the
divergences of the theory are subtracted away renormalizing $\lambda $, $a$, 
$b^{\prime }$, the Newton constant and the cosmological constant, together
with field redefinitions. The causality violations are due to the
resummation of derivatives in the vertices that couple matter with gravity.

The analysis of causality violations has been performed in a regime in which
the gravitational field is weak, which means much smaller than the Planck
mass, but rapidly varying. For a gravitational field of the order of the
Planck mass or higher it is necessary to treat the Einstein equations
coupled with matter exactly or with more powerful approximation methods. In
principle there might exist causal strong-field configurations. Here it was
important to show that there do exist configurations that violate causality
at small distances.

The causality violations are governed by the parameters $a$ and $b^{\prime }$%
. Their values need to be experimentally measured. Bounds can be derived
from the tests on the validity of Newton's law at short distances. At the
tree level, causal prescriptions exist, if $a$ and $b^{\prime }$ are
negative, but the radiative corrections make $a$ and $b^{\prime }$ run and
switch their signs. Thus there always exist configurations that violate
causality at sufficiently short distances.

On the higher-derivative side of the map $\mathcal{M}$, $a$ and $b^{\prime }$
multiply combinations of the terms $R_{\mu \nu }R^{\mu \nu }$ and $R^{2}$.
The map $\mathcal{M}$ provides a new interpretation of the physical meaning
of such terms.

Strictly speaking, the investigation of this paper makes sense only if
gravity is ultimately classical in nature. More generally, the knowledge
provided by this research may be interesting to suggest experiments to
decide whether gravity must be quantized or not.

Although the map $\mathcal{M}$ does not generalize straightforwardly to
quantum gravity, some conclusions of this paper could. Quantum gravity,
being non-renormalizable, is necessarily non-polynomial in the fields and
their derivatives. Quite generally, the resummation of derivatives can
produce causality violations, with a mechanism similar to the one
illustrated here. Therefore, it is reasonable to expect that high-energy
causality violations take place also in quantum gravity.

\vskip 25truept \noindent {\Large \textbf{Acknowledgments}}

\vskip 15truept \noindent

I am grateful to M. Halat for discussions and especially M. Mintchev and G.
Morchio for help with the generalized functions. I would also like to thank
the referee for pointing out relevant literature on approaches different
from the ones pursued here.


\begin{thebibliography}{99}
\bibitem{bohr}  N. Bohr and S. Rosenfeld, Zur frage der messbarkeit der
electromagnetischen feldgrossen, Kgl. Dansk Vidensk. Selsk, Math.-fys. Medd.
12 (1933)\ 8.

\bibitem{rosenfeld}  S. Rosenfeld, On quantization of fields, Nucl. Phys. 40
(1963) 353.

\bibitem{feynman}  R.P. Feynman, F.B. Moringo and W.G. Wagner, \textit{%
Feynman lectures on gravitation}, Penguin Books ltd, 2000.

\bibitem{moller}  C. M\o ller, in \textit{Les theories relativistes de la
gravitation}, edited by A. Lichnerowics and M.A. Tonnelat, CNRS, Paris, 1962.

\bibitem{eppley}  K. Eppley and E. Hannah, The necessity of quantizing the
gravitational field, Found. Phys. 7 (1977) 51.

\bibitem{mattingly}  J. Mattingly, Why Eppley and Hannah's experiment isn't,
arXiv:gr-qc/0601127.

\bibitem{peskin}  See for example, M. E. Peskin and D.V. Shroeder, \textit{%
An introduction to Quantum field theory}, Westview Press, 1995, chap. 20.

\bibitem{birreldavis}  B. DeWitt, \textit{Dynamical theory of groups and
fields}, Gordon And Beach Science Publishers, New York, 1965;

N.D. Birrel and P.C.W. Davies, \textit{Quantum fields in curved space},
Cambridge University Press, Cambridge, 1982.

\bibitem{schwingerkeldysh}  J. Schwinger,\textit{\ }Brownian motion of a
quantum oscillator, J. Math. Phys. 2 (1961) 407;

L.V. Keldysh, Diagram technique for nonequilibrium processes, Sov. Phys.
JETP 20 (1965) 1018.

\bibitem{magnano}  G. Magnano and L. Sokolowski, Nonlinear massive spin-2
field generated by higher derivative gravity, Ann. Phys. 306 (2003) 1.

\bibitem{jackson}  J.D. Jackson, \textit{Classical electrodynamics}, John
Wiley and Sons, Inc. (1975), chap. 17.

\bibitem{bel}  L. Bel and H. Sirouss Zia, Regular reduction of relativistic
theories of gravitation with a quadratic lagrangian, Phys. Rev. D 32 (1985)
3128.

\bibitem{parkersimon}  L. Parker and J.Z. Simon, Einstein equation with
quantum corrections reduced to second order, Phys. Rev. D 47 (1993) 1339 and
arXiv:gr-qc/9211002.

\bibitem{mottola}  See for example P.R. Anderson, C. Molina-Par\`{i}s and E.
Mottola, Linear response, validity of semi-classical gravity and the
stability of flat space, Phys. Rev. D 67 (2003) 024026 and
arXiv:gr-qc/0209075, and references therein.

\bibitem{jordan}  R.D. Jordan, Effective field equations for expectation
values, Phys. Rev. D. 33 (1986) 444.

\bibitem{woodard}  L.H. Ford and R.P. Woodard, Stress tensor correlators in
the Schwinger-Keldysh formalism, Class. Quant. Grav. 22 (2005) 1637 and
arXiv:gr-qc/0411003.

\bibitem{thooftveltman}  G. 't Hooft and M.J. Veltman, \textit{Diagrammar},
CERN report 1973-009, eq. (6.17), available at
http://preprints.cern.ch/cgi-bin/setlink?base=cernrep\&categ=Yellow\_Report%
\&id=1973-009

\bibitem{back}  B.S. DeWitt, Quantum theory of gravity. II. The manifestly
covariant theory, Phys. Rev. 162 (1967) 1195;

L.F. Abbott, The background field method beyond one loop, Nucl. Phys. B 185
(1981) 189.

\bibitem{renscal}  D. Anselmi, Renormalization of a class of
non-renormalizable theories, JHEP 07 (2005) 077 and arXiv:hep-th/0502237.

\bibitem{gravitytests}  C.D. Hoyle, B.R. Heckel, E.G. Adelberger, J.H.
Gundlach, U. Schmidt and H.E. Swanson, Sub-millimeter tests of the
gravitational inverse-square law, arXiv:hep-ph/0405262.

\bibitem{nonre}  D. Anselmi, Finiteness of quantum gravity coupled with
matter in three spacetime dimensions, Nucl. Phys. B 687 (2004) 124 and
hep-th/0309250;

D. Anselmi, Consistent irrelevant deformations of interacting conformal
field theories, JHEP 10 (2003) 045 and hep-th/0309251;

D. Anselmi, Renormalization of a class of non-renormalizable theories, JHEP\
07 (2005) 077 and hep-th/0502237;

D. Anselmi, Infinite reduction of couplings in non-renormalizable quantum
field theory, JHEP 08 (2005) 029 and hep-th/0503131;

D. Anselmi and M. Halat, Dimensionally continued infinite reduction of
couplings, JHEP 01 (2006) 077 and hep-th/0509196.

\bibitem{abse}  D. Anselmi, Absence of higher derivatives in the
renormalization of propagators in quantum field theories with infinitely
many couplings, Class. Quantum Grav. 20 (2003) 2355 and arXiv:hep-th/0212013.

\bibitem{vannieu}  P. van Nieuwenhuizen, On the renormalization of quantum
gravitation without matter, Ann. Phys. 104 (1977) 197;

P. Gilkey, The spectral geometry of a Riemannian manifold, J. Diff. Geom. 10
(1975) 601.

\bibitem{3dgrav}  D. Anselmi, Renormalization of quantum gravity coupled
with matter in three dimensions, Nucl. Phys. B 687 (2004) 143 and
hep-th/0309249.

\bibitem{gelfand}  I.M. Gel'fand and G.E. Shilov, \textit{Generalized
functions}, Vol. I, Academic Press, New York, 1964.

\bibitem{stelle}  K.S. Stelle, Renormalization of higher derivative quantum
gravity, Phys. Rev. D 16 (1977) 953.

\bibitem{weinberg}  S. Weinberg, \textit{The quantum theory of fields},
Cambridge University Press, 2000, Vol. 1, \S\ 10.2.

\bibitem{hathrell}  S.J. Hathrell, Trace anomalies and $\lambda \varphi ^{4}$
theory in curved space, Ann. Phys. (NY) 139 (1982) 136;

S.J. Hathrell, Trace anomalies and QED\ in curved space, Ann. Phys. (NY) 142
(1982) 34;

M.D. Freeman, Renormalization of non-Abelian gauge theories in curved
space-time, Ann. Phys. (NY) 153 (1984) 339.

\bibitem{fradkin}  E.S. Fradkin and A.A. Tseytlin, Renormalizable
asymptotically free quantum theory of gravity, Nucl. Phys. B 201 (1982) 469.

\bibitem{centralcharges}  D. Anselmi, Sum rules for trace anomalies and
irreversibility of the renormalization-group flow, Acta Phys. Slov. 52
(2002) 573 and hep-th/0205039;

D. Anselmi, D.Z. Freedman, M.T Grisaru and A. Johansen, Nonperturbative
formulas for central functions in supersymmetric gauge theories, Nucl. Phys.
B 526 (1998) 543 and hep-th/9708042.
\end{thebibliography}
\end{document}